\documentclass[seceq]{ptptex}

\usepackage{graphicx}
\usepackage{bm}

\notypesetlogo                       
\preprintnumber[3cm]{
OU-TAP-256\\YITP-05-17\\RESCEU-5/05}

\title{
Constraining Cosmological Parameters by the Cosmic Inversion Method
}

\author{
Noriyuki \textsc{Kogo}$^{1,2,}$\footnote{E-mail: kogo@yukawa.kyoto-u.ac.jp}, 
Misao \textsc{Sasaki}$^{2,}$\footnote{E-mail: misao@yukawa.kyoto-u.ac.jp}
and Jun'ichi \textsc{Yokoyama}$^{3,}$\footnote{E-mail: yokoyama@resceu.s.u-tokyo.ac.jp}
}

\inst{
$^1$Department of Earth and Space Science, Graduate School of Science, \\
    Osaka University, Toyonaka 560-0043, Japan \\
$^2$Yukawa Institute for Theoretical Physics, Kyoto University, \\
    Kyoto 606-8502, Japan \\
$^3$Research Center for the Early Universe, School of Science, \\
    The Univresity of Tokyo, Tokyo 113-0033, Japan
}

\abst{
We investigate the question of 
how tightly we can constrain the cosmological parameters 
by using the ``cosmic inversion'' method 
in which we directly reconstruct the power spectrum 
of primordial curvature perturbations, $P(k)$, 
from the temperature and polarization spectra 
of the cosmic microwave background (CMB). 
In a previous paper, 
we suggested that it may be possible to constrain the cosmological parameters 
using the fact that the reconstructed $P(k)$ does not depend on 
how many polarization data we incorporate in our inversion procedure 
if and only if the correct values of the cosmological parameters are used. 
The advantage of this approach is 
that we need no assumption regarding the functional form of $P(k)$. 
In this paper, we estimate typical errors 
in the determination of the cosmological parameters 
when our method is applied to the PLANCK observation. 
We investigate constraints on 
$h$, $\Omega_b$, $\Omega_m$, and $\Omega_\Lambda$ 
through Monte Carlo simulations. 
}

\begin{document}

\maketitle

\section{Introduction} \label{INTRO}

The cosmic microwave background (CMB) is one of the most powerful tools 
to determine the values of the cosmological parameters 
and examine the properties of the primordial fluctuations. 
A number of observations have been carried out 
since the Cosmic Background Explorer (COBE) observation~\cite{COBE}. 
In particular, a recent precise observation 
made using the Wilkinson Microwave Anisotropy Probe (WMAP) has confirmed that 
our universe is consistent with a spatially flat $\Lambda$CDM universe 
with Gaussian, adiabatic, and nearly scale-invariant primordial fluctuations, 
as predicted by a simple slow-roll inflation 
model~\cite{WMAPBASIC,WMAPPARA,WMAPINF,WMAPGAUSS}. 

Nevertheless, it has been suggested that 
the primordial power spectrum of the curvature perturbation, $P(k)$, 
may have some nontrivial features, 
such as a lack of power on large scales, running of the spectral index, 
and oscillatory behavior of the spectrum on intermediate scales. 
Therefore, a conventional parameter-fitting method, 
in which one often assumes a simple functional form of $P(k)$, 
such as a power-law spectrum, is unsatisfactory. 
Instead we need to examine $P(k)$ directly from observations 
without any theoretical prejudice. 
For this purpose, there have been several attempts 
to reconstruct the primordial spectrum using the WMAP data using 
model-independent methods~\cite{MW03A,MW03B,MW05,BLWE03,SH04,SS03,TDS04}. 

Cosmic inversion is a method 
to reconstruct the primordial spectrum directly from CMB anisotropies. 
It was originally proposed 
by Matsumiya et al. (2002, 2003)~\cite{MSY02,MSY03}. 
We applied this method to the WMAP first-year data 
and showed that there are possible nontrivial features 
of $P(k)$~\cite{KMSY04}. 
Our method can reproduce fine features of $P(k)$ with a resolution of 
$\Delta k \simeq 3.7\times10^{-4}\,{\rm Mpc}^{-1}$, 
which roughly corresponds to $\Delta\ell \simeq 5$ 
in the angular power spectrum, $C_\ell$. 
In a previous work, 
we improved our method in such a manner that we can use 
the auto-correlations of both CMB temperature fluctuations (TT) 
and E-mode polarization (EE)~\cite{KSY04}. 
With polarization, we have shown that 
large numerical errors in the reconstructed $P(k)$ due to the singularities 
in the inversion formula, 
which correspond to zero points of the transfer functions, can be suppressed. 
As a result, we were able to reconstruct $P(k)$ with 
an error of a few percent in the ideal situation 
that observational errors do not exist. 
We have also found that 
it may be possible to constrain the cosmological parameters 
by varying the contribution of the polarization in the inversion formula 
and requiring that the resultant $P(k)$ is independent of 
the contribution of the polarization. 
In a conventional parameter-fitting method, 
it appears that one can determine the cosmological parameters 
up to a few percent from recent WMAP data. 
However, this is because the functional space of $P(k)$ is restricted 
by the assumption of a simple functional form. 
As a result, these values of the cosmological parameters 
depend on this assumption. 
On the other hand, if we regard $P(k)$ as a free function to be reconstructed, 
there remains the degeneracy that varying the shape of $P(k)$ can compensate 
for the variation of the cosmological parameters~\cite{SBKET98,WHK01}. 
Therefore, it is important to investigate 
how well we can determine the cosmological parameters 
without any assumption on the functional form of $P(k)$. 

In this paper, we examine the proposition of 
determining the cosmological parameters by using the cosmic inversion method. 
We first confirm that it is possible 
to determine the cosmological parameters quite accurately 
when there is no observational error. 
Then we add artificial observational errors assuming the PLANCK 
observation\footnote{http://www.rssd.esa.int/index.php?project=PLANCK} 
and estimate probability distributions of the cosmological parameters 
by performing Monte Carlo simulations 
for each set of the cosmological parameters. 

This paper is organized as follows. 
In \S\ref{INVERSION}, we review our cosmic inversion method 
that employs both the CMB temperature and polarization spectra. 
We also extend it to a nonflat universe 
and examine the effect of observational errors on the reconstructed $P(k)$. 
In \S\ref{PARAMETERS}, we describe a new method 
to constrain the cosmological parameters by using our cosmic inversion method 
and report the results of simulations 
to estimate the errors on the cosmological parameters. 
Finally, we present our conclusion in \S\ref{CONCLUSION}. 

\section{Inversion method} \label{INVERSION}

\subsection{Formula} \label{FORMULA}

First we present the formula to reconstruct $P(k)$ using both 
the CMB temperature and polarization spectra~\cite{MSY02,MSY03,KSY04}. 
For completeness, its derivation is described 
in Appendix~\ref{DERIVATION}. 
We consider only scalar-type perturbations 
in which B-mode polarization is absent, 
and assume Gaussian and adiabatic primordial fluctuations 
throughout this paper. 

The angular power spectrum of the CMB anisotropy, $C^{X\bar{X}}_\ell$ 
[where $X$ and $\bar{X}$ are either 
the temperature fluctuations ($T$) or the E-mode polarization ($E$)], 
and the primordial power spectrum of the curvature perturbation, $P(k)$, 
are related as 
\begin{eqnarray}
C^{X\bar{X}}_\ell
=\frac{2}{\pi} \int_0^\infty \! \frac{dk}{k} \,
 k^3P(k) \, K^{X\bar{X}}_\ell(\eta_0,k),
\label{CLPK}
\end{eqnarray}
where $K^{X\bar{X}}_\ell(\eta_0,k)$ 
is the kernel specified by the Boltzmann equation. 
As we see, this is an integral equation for $P(k)$. 
To solve it, we first tentatively adopt the following two approximations. 
One is the thin last scattering surface (LSS) approximation, 
in which we perform the time integration of the transfer functions 
within the thickness of the LSS. 
The other is the small angle approximation, 
in which we introduce the new variable $r=2d \sin(\theta/2)$, 
which represents the conformal distance between two points on the LSS. 
With these assumptions, 
we obtain a first-order differential equation for $P(k)$ from the TT spectrum 
and algebraic equations for $P(k)$ from the EE and TE spectra, respectively. 
However, they all have singularities corresponding to the zero points of 
the transfer functions that relate the primordial curvature perturbation 
to the temperature and polarization multiple moments. 
The presence of these singularities leads to large numerical errors 
near them in the reconstructed $P(k)$, 
particularly when observational errors are taken into account. 

To avoid this difficulty, 
we construct a linear combination of the TT and EE formulas, 
introducing a free parameter $\alpha$, as 
\begin{eqnarray}
&&-k^2f^2(k)P'(k)+\left[ -2k^2f(k)f'(k)+kg^2(k)+\alpha kh^2(k) \right] P(k)
\nonumber \\
&&=S^{TT}(k)+\alpha S^{EE}(k),
\label{INVERSIONCOM}
\end{eqnarray}
where $f(k)$, $g(k)$, and $h(k)$ are time-integrated transfer functions 
within the thickness of the LSS, and the source functions are defined by 
\begin{eqnarray}
S^{TT}(k) &\equiv& 
4\pi \int_0^\infty \! dr \, \frac{1}{r} \frac{\partial}{\partial r} 
\left\{ r^3 \sum_{\ell=\ell_{\rm min}}^{\ell_{\rm max}} \frac{2\ell+1}{4\pi}
\frac{C^{TT,\,{\rm obs}}_\ell}{b^{TT,\,(0)}_\ell}
P_\ell \left( 1-\frac{r^2}{2d^2} \right) \right\} \sin kr,
\label{SKTT} \\
S^{EE}(k) &\equiv& 
4\pi \int_0^\infty \! dr \, r
\sum_{\ell=\ell_{\rm min}}^{\ell_{\rm max}}
\frac{2\ell+1}{4\pi} \frac{(\ell-2)!}{(\ell+2)!} \, 
\frac{C^{EE,\,{\rm obs}}_\ell}{b^{EE,\,(0)}_\ell}
P_\ell \left( 1-\frac{r^2}{2d^2} \right) \sin kr.
\label{SKEE}
\end{eqnarray}
Here, $C^{XX,\,{\rm obs}}$ is the observed spectrum, and 
$b^{XX,\,(0)}_\ell\equiv C^{XX,\,{\rm ex}(0)}_\ell/C^{XX,\,{\rm app}(0)}_\ell$ 
is the ratio of the exact spectrum to an approximated spectrum 
calculated from a fiducial spectrum $P^{(0)}(k)$, 
such as a scale-invariant one. 
This ratio $b^{XX}_\ell$ turns out to be almost independent of $P(k)$, 
and it therefore plays the role of a corrector of 
the errors caused by the approximations. 
The boundary conditions are given by the values of $P(k)$ 
at the zero points of $f(k)$ as 
\begin{eqnarray}
P(k_s)=\frac{S^{TT}(k_s)+\alpha S^{EE}(k_s)}
            {k_s \left[ g^2(k_s)+\alpha h^2(k_s) \right]}
\quad {\rm for} \quad f(k_s)=0,
\label{PKCOM}
\end{eqnarray}
assuming that $P'(k)$ is finite at $k=k_s$. 
In Eq.~(\ref{INVERSIONCOM}), 
the terms that have a prefactor of $\alpha$ come from the EE spectrum, 
and $\alpha$ controls the contribution of EE relative to TT. 
Because the positions of the singularities for TT and EE, 
given by $f(k)=0$ and $h(k)=0$, respectively, are different, 
if we choose an appropriate value of $\alpha$ 
such that the contribution of EE is comparable to that of TT, 
the solution of Eq.~(\ref{INVERSIONCOM}) becomes numerically stable, 
even in the neighborhoods of the singularities. 
We found such an appropriate value of $\alpha$ 
to be in the range $10^{13}-10^{15}$ 
if there is no observational error~\cite{KSY04}. 
As shown in a previous work, 
because of the tight coupling of the photon and baryon fluids at the LSS, 
the transfer function for EE, $h(k)$, is much smaller than those for TT, 
$f(k)$ and $g(k)$, by a factor of $\sim 10^{-7}$. 
The origin of the smallness of $h(k)$ is briefly explained 
in Appendix~\ref{DERIVATION}. 
This is why the appropriate value of $\alpha$ is 
in the range $10^{13}-10^{15}$. 

Our original formalism described in Appendix~\ref{DERIVATION} is based on 
a flat universe, where the spherical Bessel functions appear as 
the radial eigenfunctions of a Fourier expansion. 
Here, we extend our method to a nonflat universe. 
The dominant effect due to the curvature of the 3-geometry 
is effectively absorbed by adjusting the conformal distance to the LSS 
so that the angular diameter distance in a curved background 
is properly recovered. 
This gives rise to a shift of the Doppler peaks, 
with the overall shape of $C^{XX}_\ell$ unchanged. 
The shape of $C^{XX}_\ell$ changes mainly on large scales (at small $\ell$), 
and its change on small scales is only a few percent~\cite{HW96,ZSB98}. 
Thus, under the small angle approximation, 
we need only modify $d$ in $C^{XX,\,{\rm app}}_\ell$ 
depending on the curvature. 
This means that we can still use the spherical Bessel functions instead of 
the ultraspherical Bessel functions, which appear as the radial eigenfunctions 
in the curved background in $C^{XX,\,{\rm app}}_\ell$. 
In fact, we find that using the spherical Bessel functions 
causes only a small error that can be corrected by $b^{XX}_\ell$ 
and does not affect the resultant $P(k)$. 

In conclusion, for a nonflat universe, 
we need no modification of our inversion formalism 
described in Appendix~\ref{DERIVATION}, except for the adjustment of $d$, 
and the inversion formula~(\ref{INVERSIONCOM}) is applicable 
regardless of the geometry.

\subsection{Effect of observational errors} \label{ERRORS}

Although in a previous paper~\cite{KSY04} we showed that 
our method is effective if we assume that there is no observational error, 
it is also important to elucidate the effect of observational errors 
on the reconstructed $P(k)$. 
The existence of observational errors causes 
numerical errors to be amplified near the singularities, 
as shown in our recent work~\cite{KMSY04}, where we used only the TT spectrum. 
Therefore, we reconstructed $P(k)$ from $C^{XX}_\ell$ 
with observational errors, 
varying $\alpha$, which represents the contribution of EE. 
We used the PLANCK observation and estimated the observational errors, 
$\Delta C^{XX}_\ell$, by using the analytic formula~\cite{LK95} 
\begin{eqnarray}
\Delta C^{XX}_\ell
=\sqrt{\frac{2}{(2\ell+1)f_{\rm sky}}}
 \left( C^{XX,\,{\rm real}}_\ell
 +4\pi\frac{\sigma_{\rm pix}^2}{N_{\rm pix}}
  e^{\ell^2 \sigma_{\rm beam}^2} \right),
\label{DELTACL}
\end{eqnarray}
where $f_{\rm sky}$ is the sky coverage 
($f_{\rm sky}$=1 for a full-sky survey), 
$N_{\rm pix}$ is the number of pixels, 
$\sigma_{\rm pix}$ is the noise per pixel, 
$\sigma_{\rm beam}$ is the beam size, 
and $C^{XX,\,{\rm real}}_\ell$ is the real power spectrum. 
We set $f_{\rm sky}=1$, $N_{\rm pix}=2.3 \times 10^6$, 
$\sigma_{\rm pix}=2.0\,{\rm \mu K}$ and $3.7\,{\rm \mu K}$ 
for TT and EE, respectively, 
and $\sigma_{\rm beam}=0.13^{\circ}$ for the PLANCK 143GHz channel. 

First, for each $\ell$, we drew a random number from a Gaussian distribution 
with $\Delta C^{XX}_\ell$ around a theoretical $C^{XX}_\ell$, 
and then reconstructed $P(k)$ from each simulated data set. 
Next, we estimated the mean and variance of 
the reconstructed $P(k)$ at each $k$ for 1000 realizations. 
Our numerical calculations for evaluating $C^{XX}_\ell$ 
and the transfer functions are based on 
CMBFAST\footnote{http://www.cmbfast.org/}. 
We used $C^{XX}_\ell$ up to $\ell_{\rm max}=1500$ for both TT and EE. 
We assume a scale-invariant spectrum, $k^3P(k)=\mbox{const.}$, 
and set the cosmological parameters as 
$h=0.70$, $\Omega_b=0.050$, $\Omega_m=0.30$, $\Omega_\Lambda=0.70$, 
and $\tau=0.20$. 
In this case, the positions of the TT singularities are at 
$kd \simeq 70$, $430$, $680$, $1030$, $\cdots$, 
and those of the EE singularities are at 
$kd \simeq 230$, $560$, $860$, $1180$, $\cdots$, 
where $d \simeq 1.36\times10^4\,{\rm Mpc}$. 
The range of the reconstructed $P(k)$ is 
between the first and fourth TT singularities, i.e., $70 \le kd \le 1030$, 
where $kd$ roughly corresponds to $\ell$. 
To focus on the effect of observational errors, 
we assume that the cosmological parameters are known precisely. 

\begin{figure}[htb]
\begin{center}
\includegraphics[width=8cm]{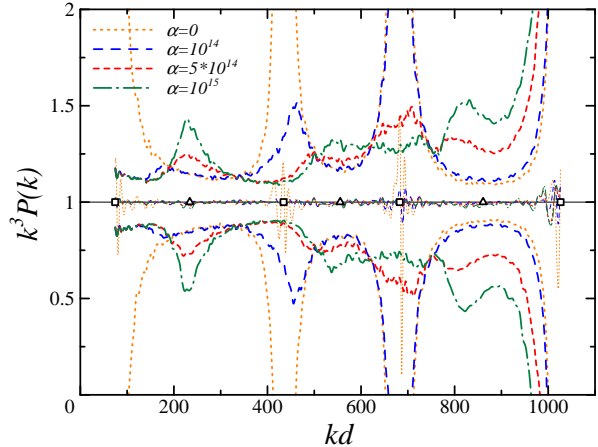}
\caption{Reconstructed spectra with observational errors. 
We assume a scale-invariant $P(k)$ and exactly known cosmological parameters, 
and we add the PLANCK observational errors 
to the theoretical $C^{TT}_\ell$ and $C^{EE}_\ell$. 
The thin and thick curves represent the mean and variance, 
obtained from 1000 simulations, respectively. 
We plot them for four different values of $\alpha$ 
($0$, $10^{14}$, $5\times10^{14}$, and $10^{15}$)
which controls the contribution of EE relative to TT. 
The horizontal axis, $kd$, roughly corresponds to $\ell$. 
The TT and EE singularities 
are represented by the symbols $\square$ and $\triangle$, respectively. 
\label{ERRE}}
\end{center}
\end{figure}

Figure~\ref{ERRE} shows the results for 
$\alpha=0$, $10^{14}$, $5\times10^{14}$, and $10^{15}$. 
We see that the error in $P(k)$ is very large near the TT singularities 
for $\alpha=0$, which is also the case in the situation studied 
in our recent work~\cite{KMSY04}. 
However, as $\alpha$ is increased, 
near the TT singularities the error is reduced, 
while near the EE singularities it is amplified. 
For $\alpha=5\times10^{14}$, the numerical error 
seems to be strongly suppressed near both the TT and EE singularities. 
On small scales (for $\ell \gtrsim 700$), 
the overall error in $P(k)$ becomes larger as $\alpha$ is increased, 
because the detector noise in $C^{EE}_\ell$ becomes 
the dominant source of error for the PLANCK observation. 
In the absence of observational errors, 
we showed that $P(k)$ is accurately reconstructed 
in the range $10^{13} \lesssim \alpha \lesssim 10^{15}$~\cite{KSY04}. 
Taking the PLANCK observational errors into account,
this range is narrowed due to the amplification of the numerical error, 
and the appropriate value of $\alpha$ is found to be $\sim 5\times10^{14}$, 
for which we can still suppress the numerical errors. 

\section{Constraining cosmological parameters} \label{PARAMETERS}

\subsection{Method} \label{METHOD}

As mentioned in a previous paper~\cite{KSY04}, 
it is in principle possible to constrain the cosmological parameters 
in our reconstruction method as follows. 
We need to examine the behavior of the reconstructed $P(k)$ 
when the free parameter $\alpha$ 
introduced in Eq.~(\ref{INVERSIONCOM}) is varied. 
We find that 
as long as we use the correct values of the cosmological parameters, 
$P(k)$ is accurately reconstructed for an appropriate value of $\alpha$. 
On the other hand, if we use incorrect values, 
we obtain a particular deformed shape of the reconstructed $P(k)$ 
that depends on the value of $\alpha$. 
To be more precise, if we use a relatively large value of $\alpha$, 
the deformation appears near the EE singularities, 
while if we use a smaller value, 
the deformation appears near the TT singularities. 
The point is that such deformations 
caused by the incorrect choice of the cosmological parameters 
indicate not only deviations from the actual $P(k)$ 
but also inconsistent results for $P(k)$ among different values of $\alpha$. 

Let us denote the reconstructed $P(k)$ for a certain value of $\alpha$ 
by $P_\alpha(k)$ and introduce a quantity that represents the difference 
between $P_{\alpha_1}(k)$ and $P_{\alpha_2}(k)$. 
We define 
\begin{eqnarray}
D \equiv \int_{k_{\rm min}}^{k_{\rm max}} \! \frac{dk}{k} \, 
\left[ k^3P_{\alpha_1}(k) - k^3P_{\alpha_2}(k) \right]^2,
\label{DEVI}
\end{eqnarray}
where $\alpha_1 \ne \alpha_2$. 
From the above argument, 
we speculate that $D$ takes its minimum value 
with respect to variation of the cosmological parameters 
at the correct values of these parameters. 
To confirm this speculation, 
we estimated the values of $D$ for 
different values of the cosmological parameters, 
assuming that the values of these parameters 
used in \S\ref{ERRORS} are the true values. 
We used $C^{XX}_\ell$ up to $\ell_{\rm max}=1000$ for both TT and EE 
to avoid the effect of a large observational error. 
Hence, we take $k_{\rm min}$ and $k_{\rm max}$ 
to be the first and third TT singularities, respectively. 
We used $\alpha_1=10^{15}$ and $\alpha_2=10^{14}$, 
because these values are appropriate 
to suppress large numerical errors due to the singularities. 

First, we assumed no observational error in $C^{XX}_\ell$ 
and estimated $D$ as a function of the cosmological parameters. 
To see the dependence of $D=D(h,\Omega_b,\Omega_m,\Omega_\Lambda)$ on $P(k)$, 
we performed calculations for three different shapes of $P(k)$. 
We adopted a scale-invariant $P(k)$, one with a large peak and dip, 
and one with a small oscillation. 
The spectrum with a peak and dip is expressed as 
\begin{eqnarray}
k^3 P(k)
=A \left\{1+a_1\exp\left[ -\frac{(k-k_1)^2}{\sigma_1{}^2} \right]\right\}
   \left\{1+a_2\exp\left[ -\frac{(k-k_2)^2}{\sigma_2{}^2} \right]\right\}^{-1},
\label{PKPD}
\end{eqnarray}
and we set $a_1=a_2=1$, 
$k_1=0.03\,{\rm Mpc}^{-1}$, $k_2=0.06\,{\rm Mpc}^{-1}$, 
$\sigma_1=0.01\,{\rm Mpc}^{-1}$, and $\sigma_2=0.005\,{\rm Mpc}^{-1}$.
The spectrum with an oscillation is expressed as 
\begin{eqnarray}
k^3 P(k)
=A \left[ 1+a_0\sin\left( \frac{k}{k_0} \right) \right],
\label{PKOSC}
\end{eqnarray}
and we set $a=0.1$ and $k_0=5\times10^{-4}\,{\rm Mpc}^{-1}$. 
To elucidate the dependence of $D$ on each cosmological parameter, 
we varied $h$, $\Omega_bh^2$, $\Omega_mh^2$, 
and $\Omega_K=1-\Omega_m-\Omega_\Lambda$ 
one at a time, keeping the others fixed to the assumed real values. 
Explicitly, we varied $h$ from $0.60$ to $0.80$ with $\Delta h=0.01$, 
keeping $\Omega_bh^2$, $\Omega_mh^2$, and $\Omega_K$ fixed; 
$\Omega_b$ from $0.040$ to $0.060$ with $\Delta\Omega_b=0.001$, 
keeping $h$, $\Omega_mh^2$, and $\Omega_K$ fixed; 
$\Omega_m$ from $0.20$ to $0.40$ with $\Delta\Omega_m=0.01$, 
keeping $h$, $\Omega_bh^2$, and $\Omega_K$ fixed; 
and $\Omega_\Lambda$ from $0.60$ to $0.80$ with $\Delta\Omega_\Lambda=0.01$, 
keeping $h$, $\Omega_bh^2$, and $\Omega_mh^2$ fixed. 
Thus, the number of grid points is 21 in each case. 
Note that we fix the optical depth $\tau$ in our analysis, 
because it changes the shape of the spectrum only on large scales. 

\begin{figure}[ht]
\begin{center}
\includegraphics[width=6cm]{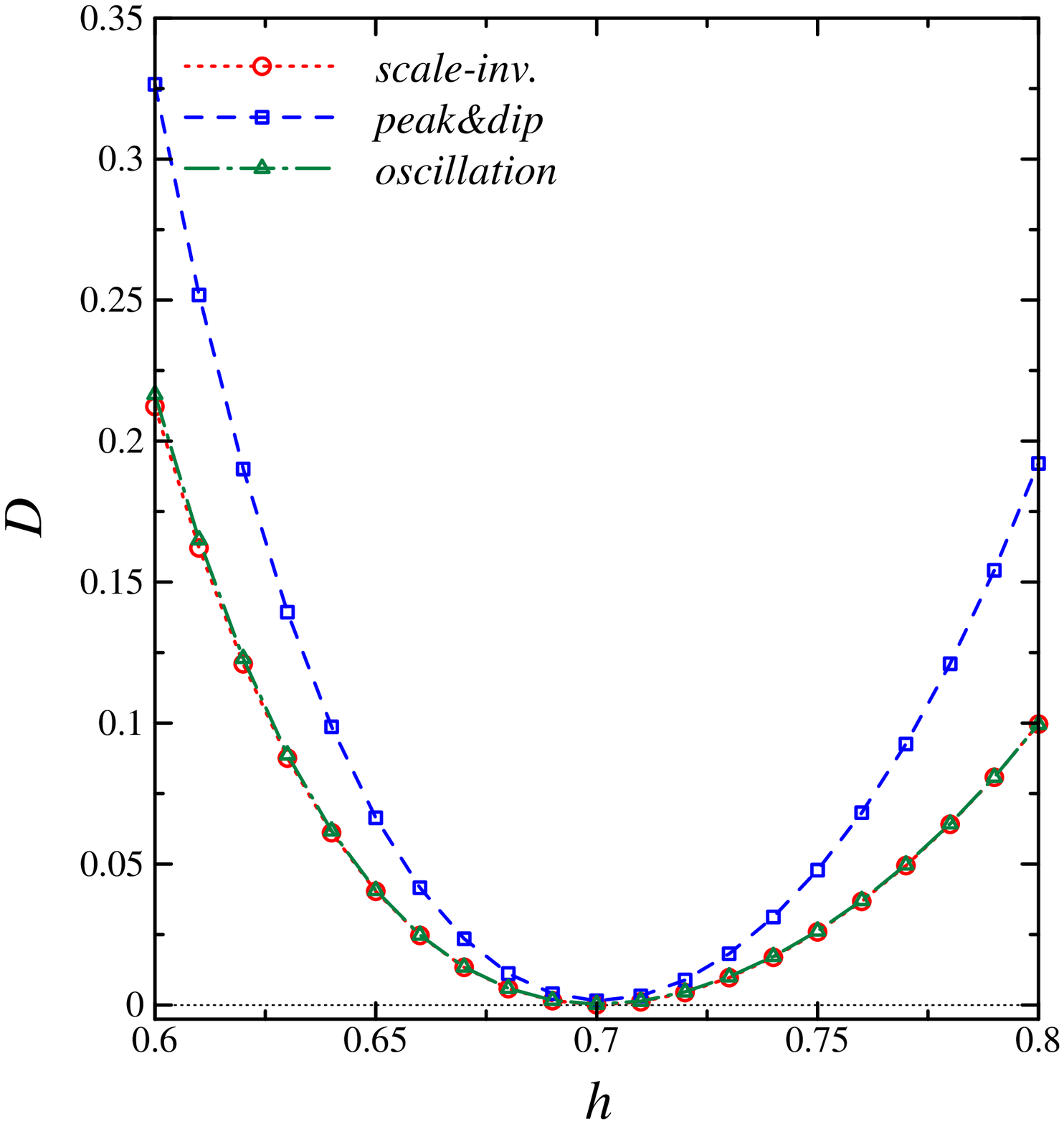}
\includegraphics[width=6cm]{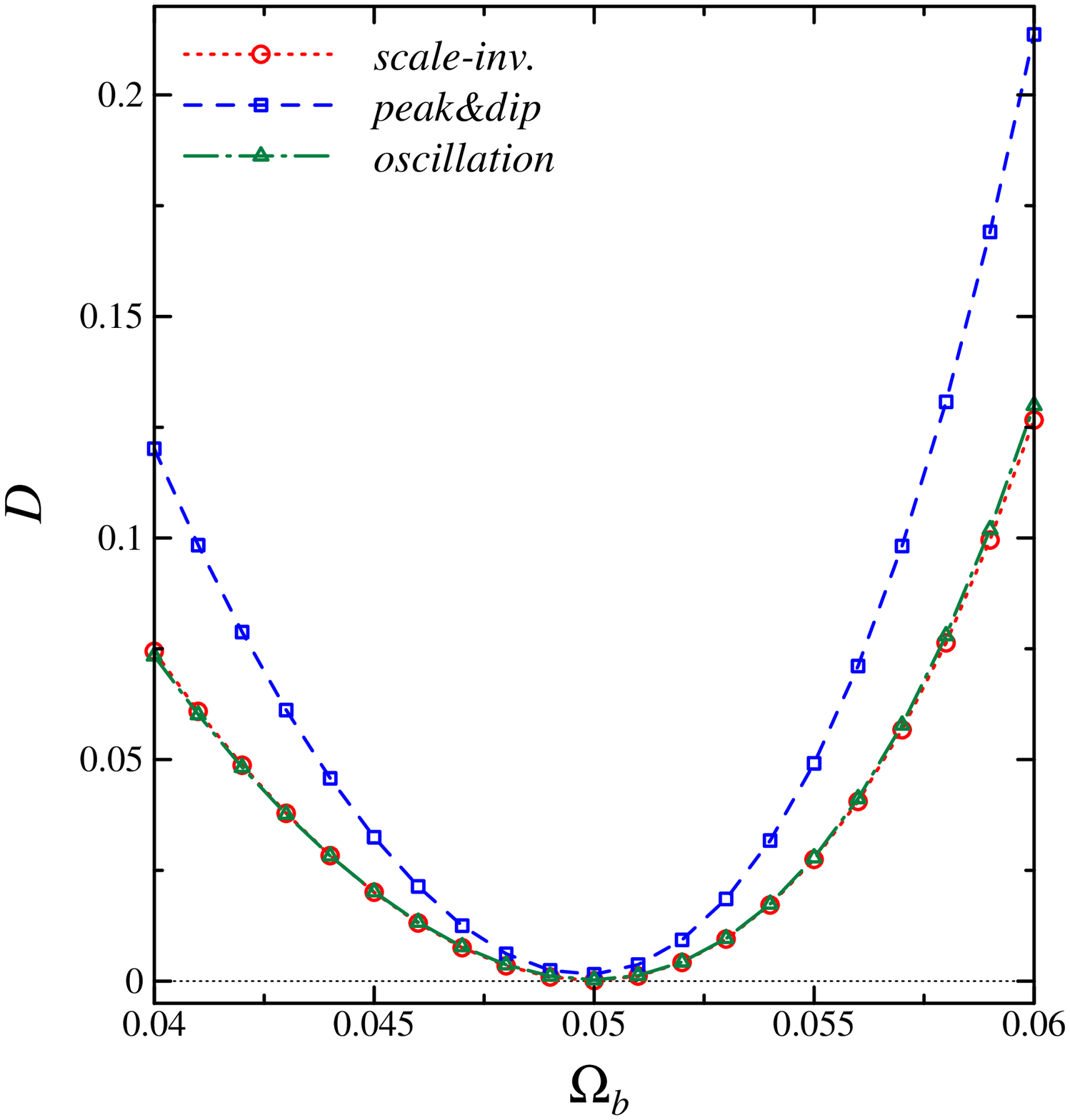}
\includegraphics[width=6cm]{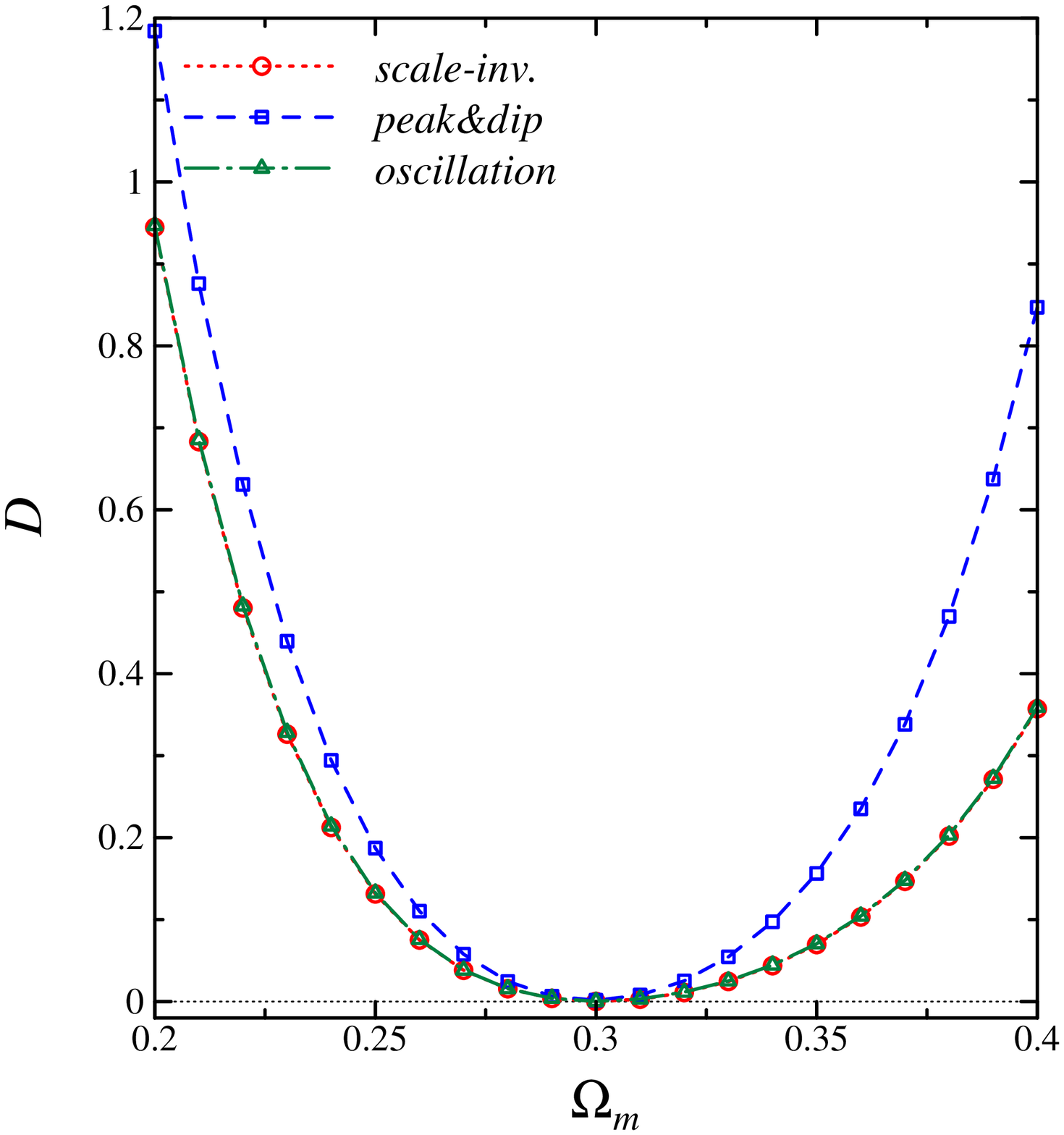}
\includegraphics[width=6cm]{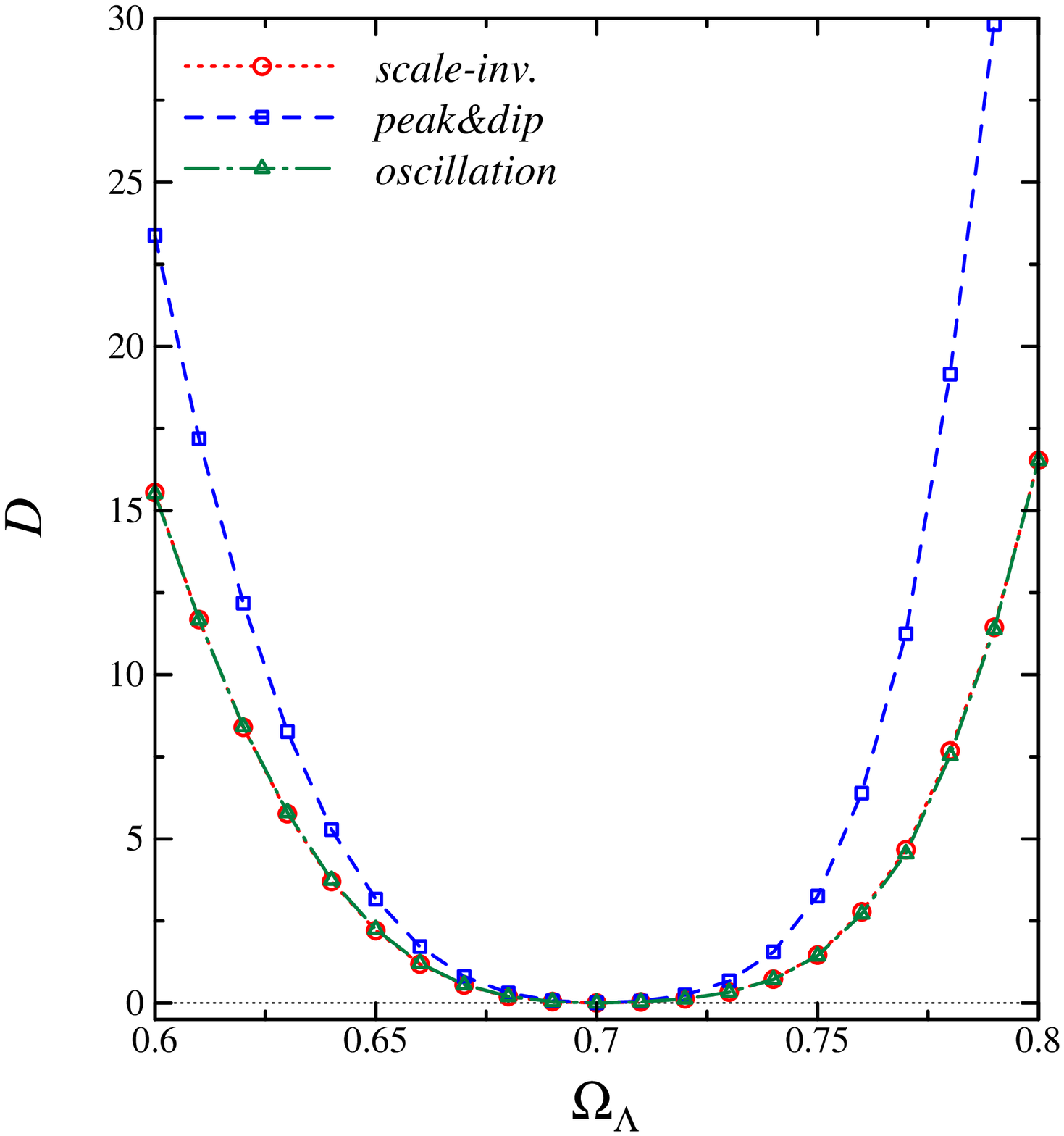}
\caption{$D$ defined by Eq.~(\ref{DEVI}) 
as a function of each cosmological parameter. 
In the separate plots in each panel, 
we assume $P(k)$ as a scale-invariant spectrum, 
a spectrum with a peak and a dip, and a spectrum with a small oscillation. 
The true cosmological parameters are assumed to be 
$h=0.70$, $\Omega_b=0.050$, $\Omega_m=0.30$, $\Omega_\Lambda=0.70$, 
and $\tau=0.20$. 
We vary $h$, $\Omega_bh^2$, $\Omega_mh^2$, and $\Omega_K$, 
keeping the others fixed at the assumed true values, 
in the top left, top right, bottom left, and bottom right panels, 
respectively. (See the text for details.) 
\label{DPARA}}
\end{center}
\end{figure}

From the results shown in Fig.~\ref{DPARA}, 
we find that regardless of the shape of $P(k)$, 
$D$ as a function of each cosmological parameter takes its minimum value 
at the correct value of that parameter in any case. 
With regard to the difference, 
the spectrum that has a large peak and dip tends to give larger values of $D$, 
while that with a small oscillation gives almost the same values 
as the scale-invariant spectrum. 
We also find that $D$ is most sensitive to $\Omega_\Lambda$ or $\Omega_K$. 
This is because the curvature affects the angular scale. 
It follows that the positions of the singularities, 
which correspond to the zero points of the transfer functions, 
shift to incorrect positions, 
and this causes large deviations of the reconstructed $P(k)$. 
As described in this section, 
we have confirmed that our method to constrain the cosmological parameters 
is effective as long as the observational error can be ignored. 
This method is quite intriguing, 
because it requires no assumption regarding the functional form of $P(k)$. 

\subsection{Error estimation} \label{ESTIMATION}

We also performed simulations 
to estimate the errors on the cosmological parameters 
in the case that we observe CMB anisotropies by the PLANCK satellite. 
We generated 1000 realizations with the PLANCK observational errors 
and reconstructed $P(k)$ from each realization in the same way 
as described in \S\ref{ERRORS}. 
For each realization, 
we calculated the value of $D$ by varying the cosmological parameters and 
finding the minimum of $D$, which represents the location of the real values 
of the cosmological parameters, as discussed in the previous subsection. 
In practice, however, the location of the real values may be 
different from the minimum of $D$, due to observational errors. 
Therefore, we constructed histograms of the values of 
the cosmological parameters at the minimum of $D$ from the 1000 realizations 
and estimated their probability distributions by Gaussian smoothing. 
The assumed model is a scale-invariant $P(k)$ with the same values of 
the cosmological parameters as used in \S\ref{ERRORS}. 
Ultimately, we should perform a wide range parameter search 
in multiple dimensions. 
However, here we used limited ranges for the possible values of the parameters 
and performed parameter searches only in one and two dimensions, 
mainly because the purpose of this paper is 
to carry out preliminary examination of how our basic strategy of 
unrestricting the functional form of $P(k)$ affects the parameter estimation, 
and, practically, because our computations are quite time consuming. 

\begin{figure}[ht]
\begin{center}
\includegraphics[width=6cm]{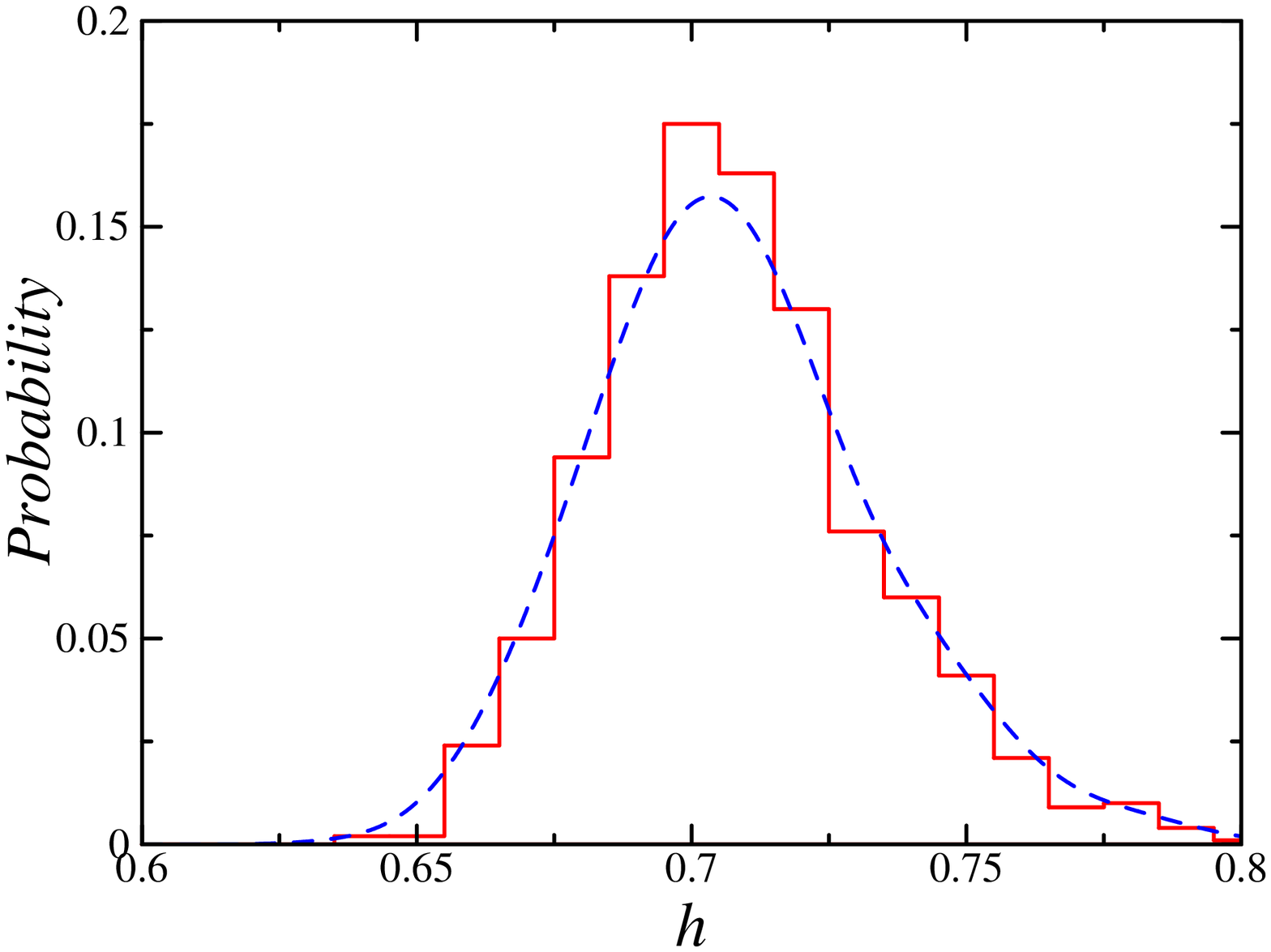}
\includegraphics[width=6cm]{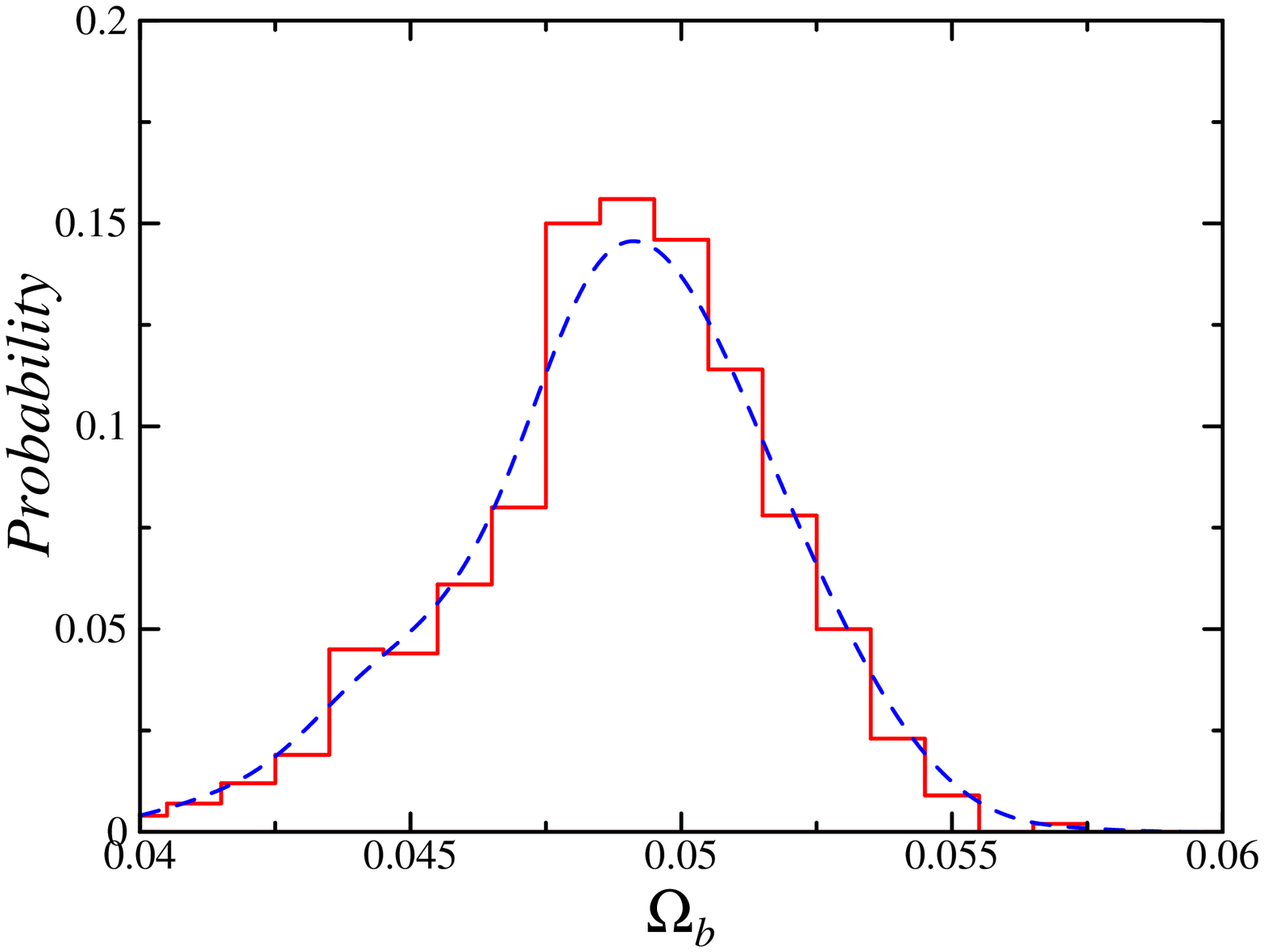}
\includegraphics[width=6cm]{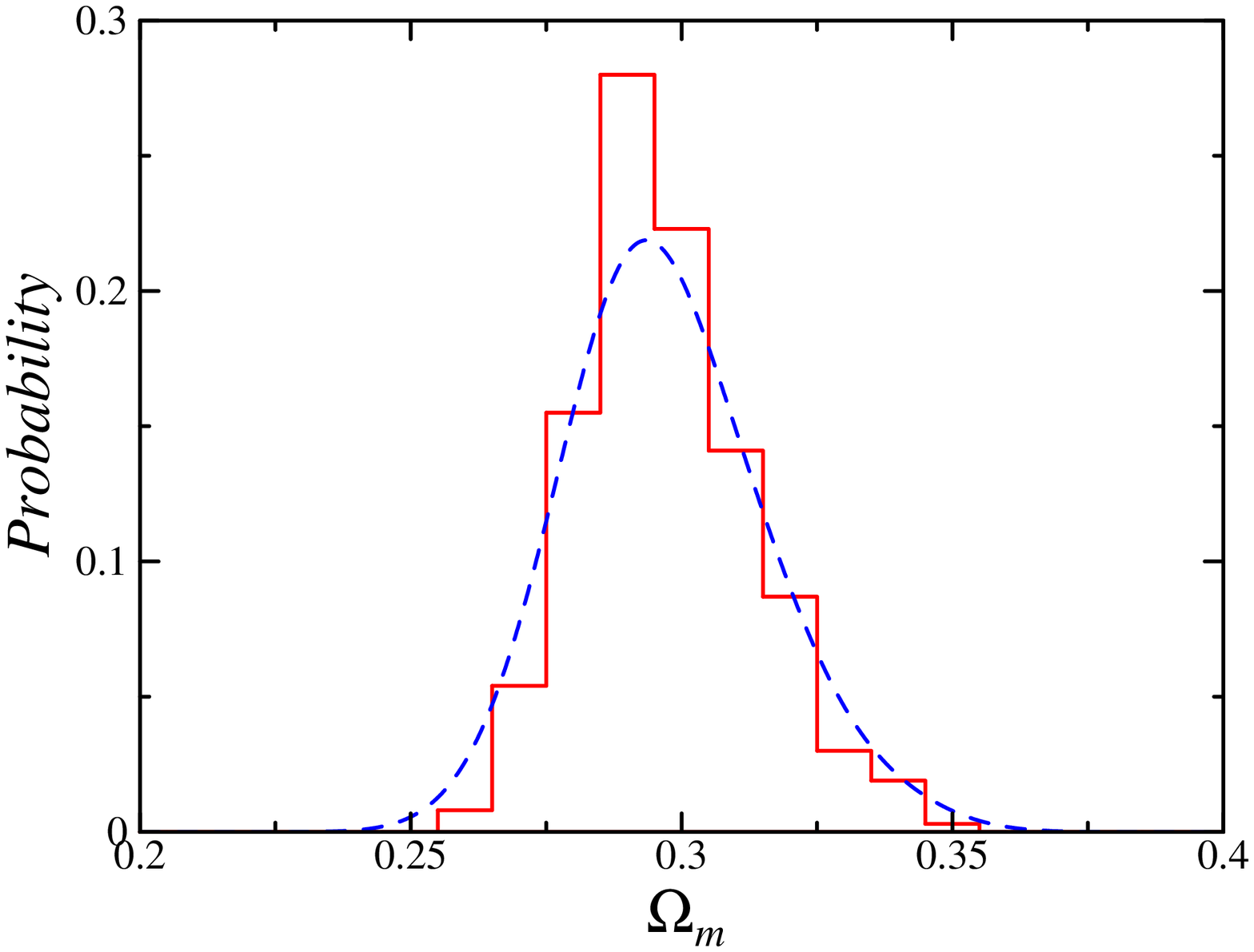}
\includegraphics[width=6cm]{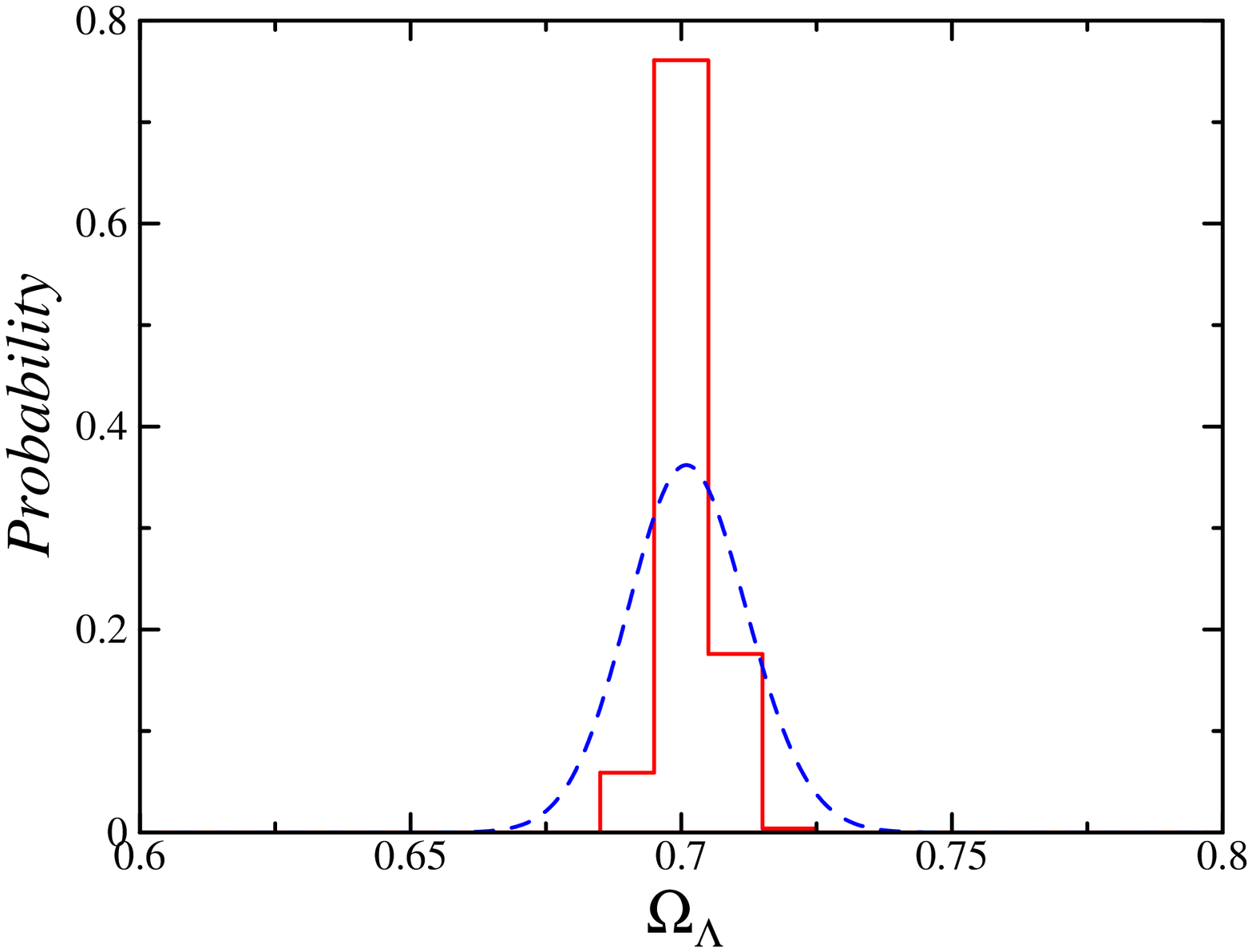}
\caption{Probability distributions of the cosmological parameters 
estimated with the one-dimensional analyses. 
The histograms were obtained from 1000 simulations, 
and the dashed curves are the estimated probability distribution functions. 
The assumed model is a scale-invariant spectrum 
with the same cosmological parameters as in Fig.~\ref{DPARA}. 
The bin sizes for the parameters are 
$\Delta h=0.01$, $\Delta\Omega_b=0.001$, $\Delta\Omega_m=0.01$, 
and $\Delta\Omega_\Lambda=0.01$. 
We vary $h$, $\Omega_bh^2$, $\Omega_mh^2$, and $\Omega_K$ individually, 
keeping the others fixed, in the top left, top right, bottom left, 
and bottom right panels, respectively. 
\label{HPARA1}}
\end{center}
\end{figure}

First, we varied each cosmological parameter individually, 
keeping the others fixed, as described in the previous subsection. 
The results are shown in Fig.~\ref{HPARA1}. 
We find that the probability distributions are nearly Gaussian, 
and their peaks lie near the correct values in all cases, as expected, 
since we drew random numbers from Gaussian distributions 
around the theoretical $C^{XX}_\ell$ in our simulations, 
as mentioned in \S\ref{ERRORS}. 
For each cosmological parameter, 
we calculated its most probable value and the $1\sigma$ error 
from its estimated probability distribution. 
The result is given in Table~\ref{VPARA1}. 
We see that the most tightly constrained parameter 
is $\Omega_\Lambda$ or $\Omega_K$, whose relative error is a few percent, 
provided that the other parameters are known and fixed. 
This is because $D$ is most sensitive to the curvature, 
as mentioned in the previous subsection. 

\begin{figure}[ht]
\begin{center}
\includegraphics[width=8cm]{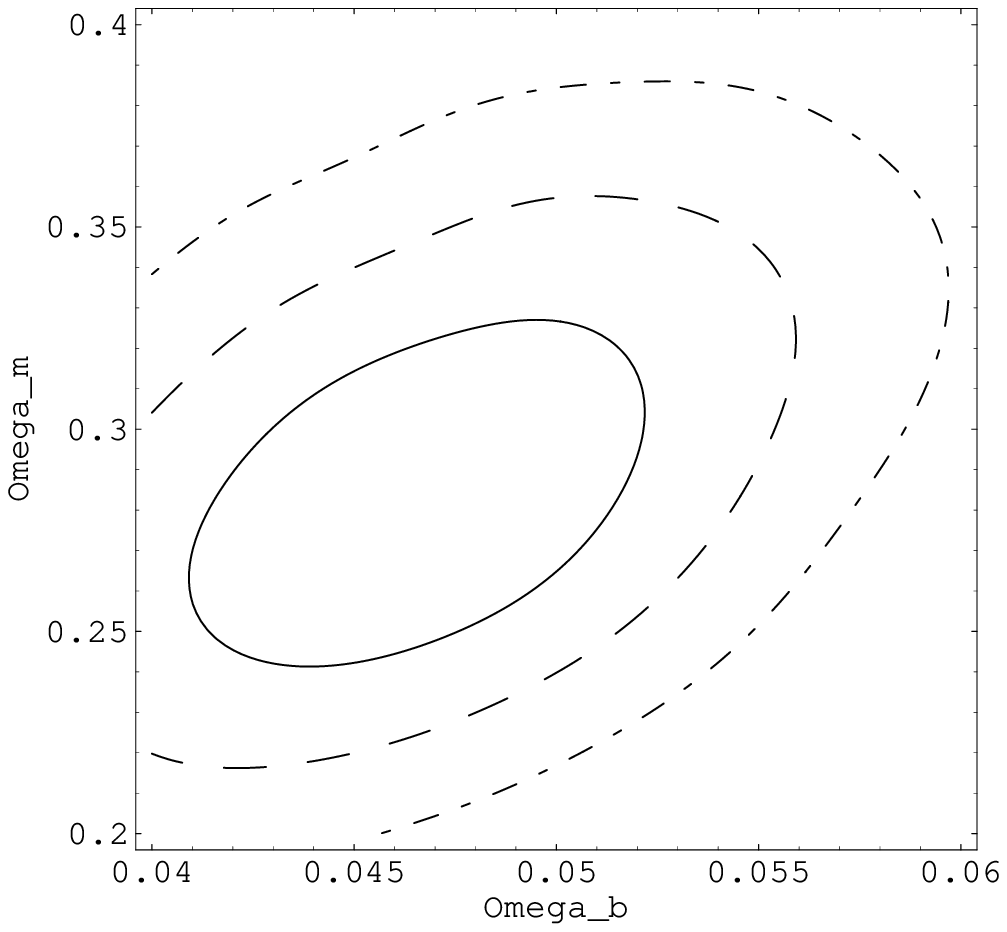}
\includegraphics[width=6cm]{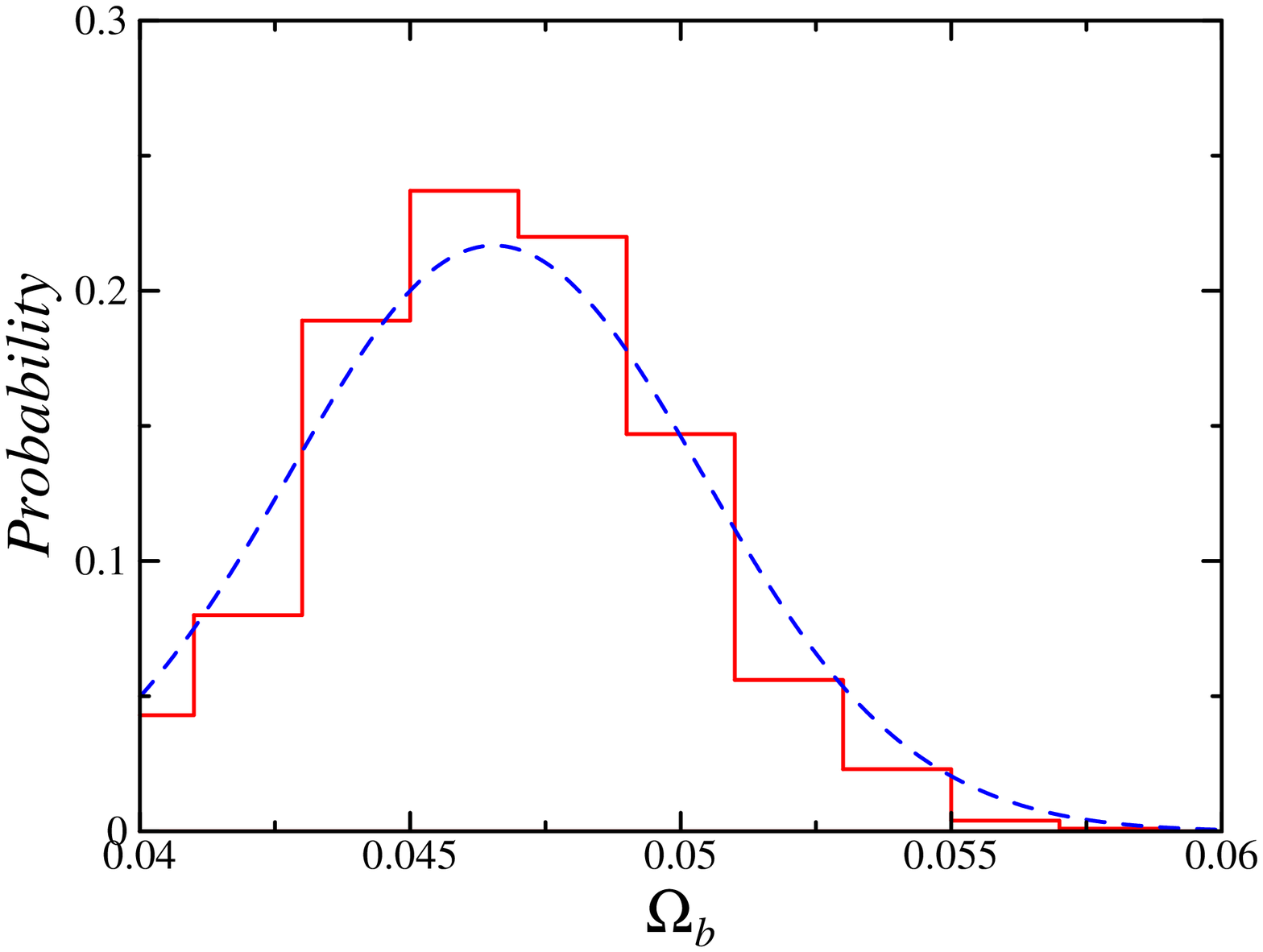}
\includegraphics[width=6cm]{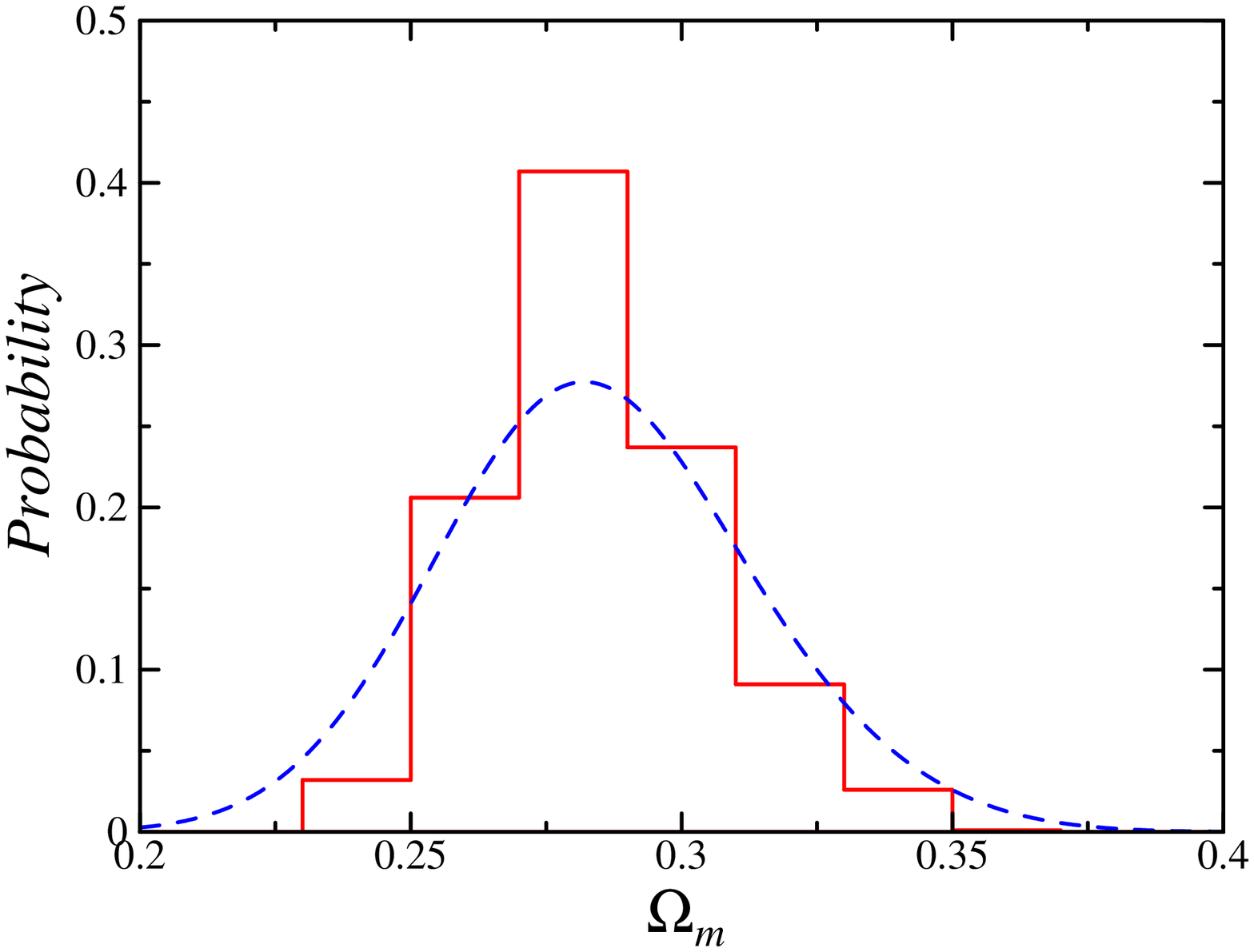}
\caption{Probability distributions of the cosmological parameters 
estimated with the two-dimensional analysis 
in $(\Omega_b\,,\, \Omega_m)$ space. 
The assumed model is the scale-invariant spectrum 
with the same cosmological parameters as in Fig.~\ref{DPARA}. 
The bin sizes for the parameters are taken as 
$\Delta\Omega_b=0.002$ and $\Delta\Omega_m=0.02$. 
The values of $h$ and $\Omega_K$ are fixed at the assumed values. 
The top panel shows the two-dimensional probability distribution, 
where the solid, dashed, and dash-dotted curves represent 
the 1$\sigma$, 2$\sigma$, and 3$\sigma$ regions, respectively, 
and the bottom left and right panels show 
the projected probability distributions 
for $\Omega_b$ and $\Omega_m$, respectively. 
\label{HBM2}}
\end{center}
\end{figure}

\begin{figure}[ht]
\begin{center}
\includegraphics[width=8cm]{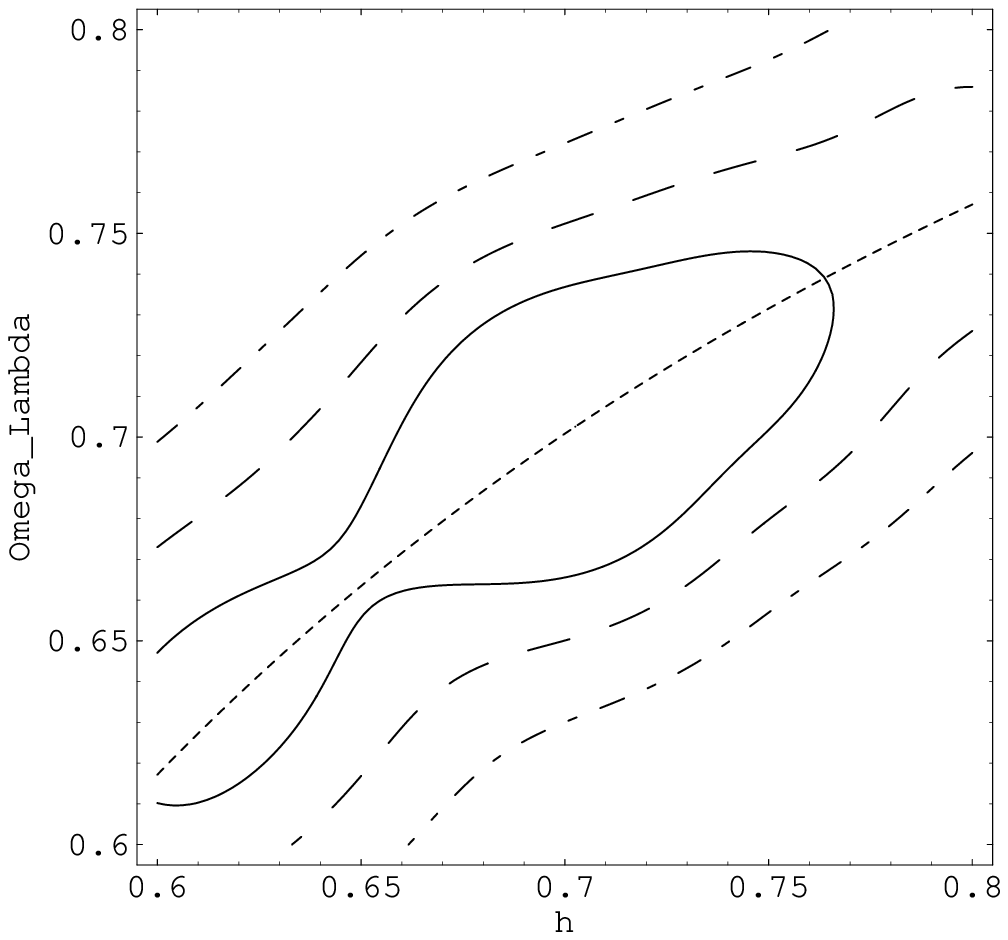}
\includegraphics[width=6cm]{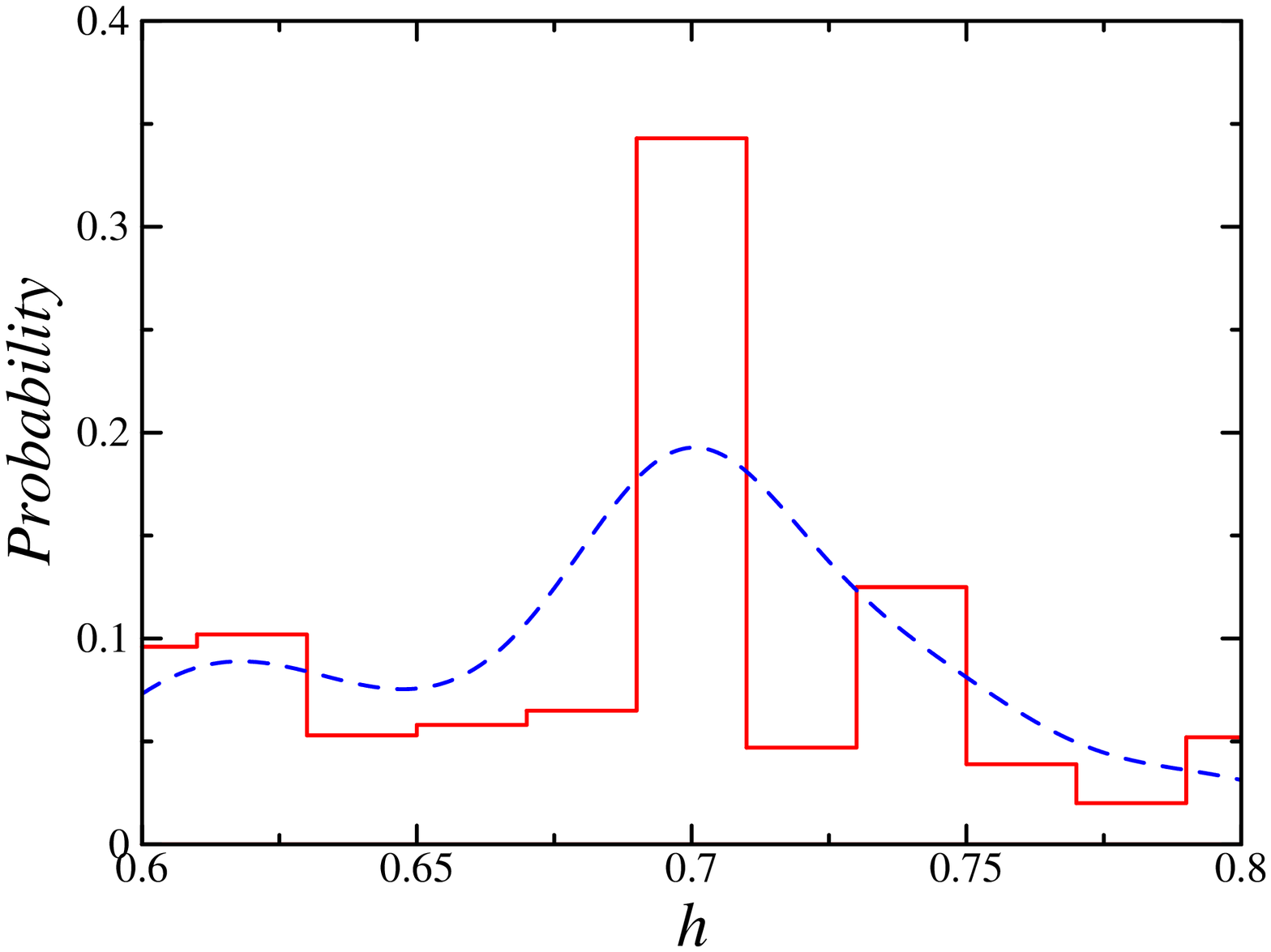}
\includegraphics[width=6cm]{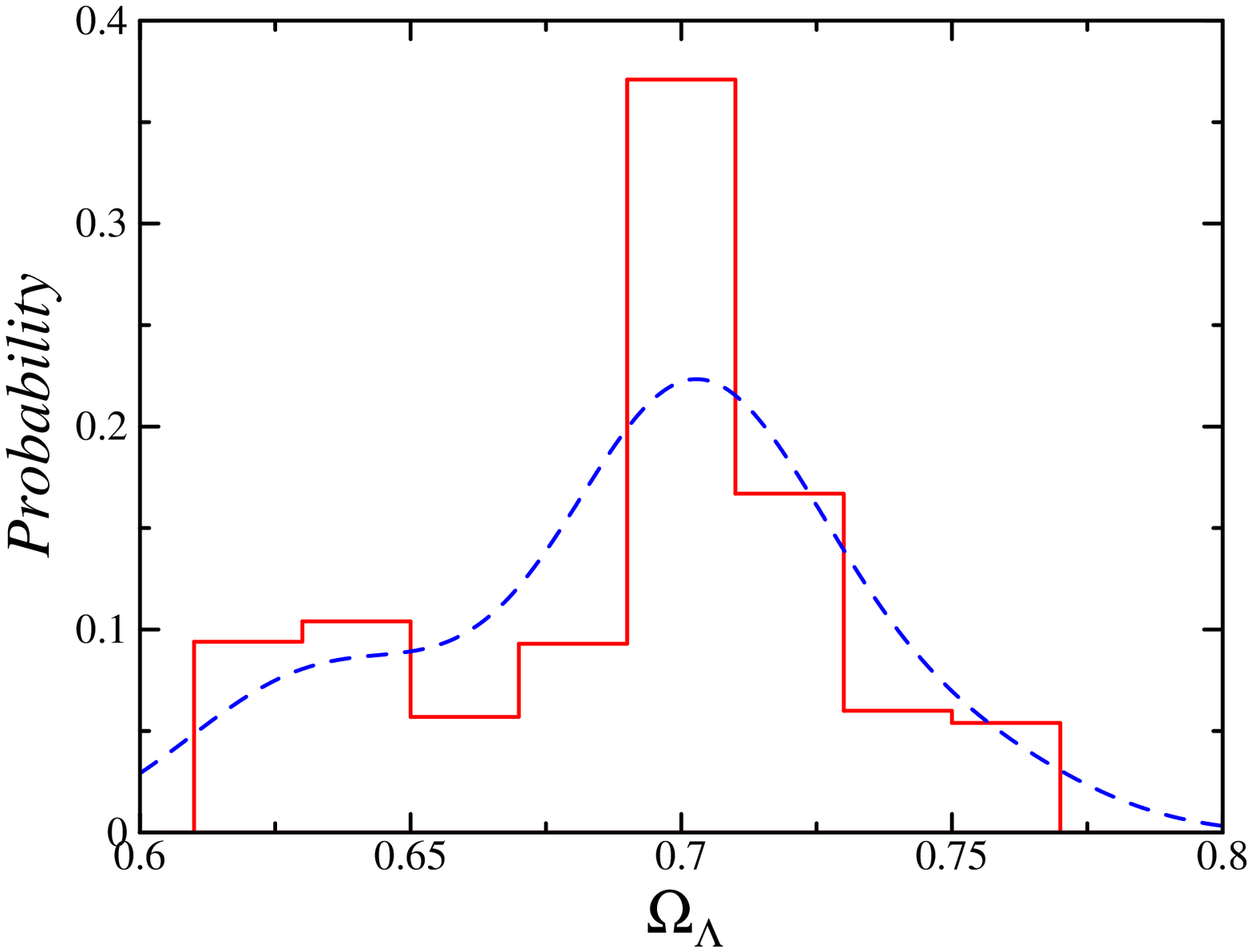}
\caption{Probability distributions of the cosmological parameters 
estimated with the two-dimensional analysis 
in $(h\,,\,\Omega_\Lambda)$ space. 
The assumed model is the scale-invariant spectrum 
with the same cosmological parameters as in Fig.~\ref{DPARA}. 
The bin sizes for the parameters are taken as 
$\Delta h=0.02$ and $\Delta\Omega_\Lambda=0.02$. 
We vary $h$ and $\Omega_\Lambda$, 
keeping $\Omega_bh^2$ and $\Omega_mh^2$ fixed. 
The top panel shows the two-dimensional probability distribution, 
where the solid, dashed, and dash-dotted curves represent 
the 1$\sigma$, 2$\sigma$, and 3$\sigma$ regions, respectively, 
and the bottom left and right panels show 
the projected probability distributions 
for $h$ and $\Omega_\Lambda$, respectively. 
The short-dashed curve in the top panel represents the degeneracy curve, 
where the angular diameter distance is the same as that of the assumed model. 
\label{HHV2}}
\end{center}
\end{figure}

To see possible degeneracies among the cosmological parameters, 
we also performed two sets of two-dimensional analyses. 
In these analyses, we varied $\Omega_b$ and $\Omega_m$, 
while keeping $h$ and $\Omega_K$ fixed, 
and varied $h$ and $\Omega_\Lambda$, 
while keeping $\Omega_bh^2$ and $\Omega_mh^2$ fixed. 
To save computational time, 
we investigated the same range of the parameter space, 
but with bin sizes twice as large as the corresponding ones 
in the one-dimensional analyses. 
Thus, the number of grid points is $11\times11$. 
The results for $\Omega_b$ and $\Omega_m$ are shown in Fig.~\ref{HBM2}, 
and the estimated values of the cosmological parameters 
are listed in Table~\ref{VBM2}. 
We find that in both cases the peaks deviate slightly 
from their correct values. 
In particular, the deviation from the correct value of $\Omega_b$ 
is quite large. 
As shown in Fig.~\ref{HHV2}, for $h$ and $\Omega_\Lambda$, 
there is a degeneracy caused by the fact that 
the same angular diameter distance can be the same 
for different sets of values of $h$ and $\Omega_\Lambda$~\cite{EB99}. 
We also see a peak near the correct values, 
but this peak is artificial, 
due to the sparseness of the parameter values we used. 
Moreover, the peaks in the projected probability distributions 
are amplified due to the narrowness of the parameter range, 
in which the global probability distribution does not converge sufficiently. 
For these reasons, we cannot estimate the values of the cosmological paramters 
from this analysis. 
We need more fine-meshed and wide-ranged parameter search, 
although the computational time becomes quite long in this case. 

\section{Conclusion} \label{CONCLUSION}

In a previous work~\cite{KSY04}, we proposed a method 
to reconstruct the primordial spectrum 
by using both the CMB temperature (TT) and the polarization (EE) spectra, 
and we showed that it is effective if there is no observational error. 
We also proposed a new method to constrain the cosmological parameters 
by using the fact that the shape of $P(k)$ obtained in this way 
must be independent of the contribution of EE relative to TT, 
which is controlled by the dimensionless parameter $\alpha$. 
We found that the resulting $P(k)$ depends very strongly on $\alpha$, 
unless the correct values of the cosmological parameters 
are used in the reconstruction. 
Using this fact, we can, in principle, constrain the cosmological parameters 
without any assumption on the functional form of $P(k)$. 

In this paper, first, to elucidate the effect of observational error, 
we have reconstructed $P(k)$ from $C_\ell$ 
with the errors expected from PLANCK satellite observations, 
assuming that the cosmological parameters are known. 
As mentioned above, 
the contribution of EE relative to TT is parameterized by $\alpha$. 
We have found that numerical errors due to the singularities 
corresponding to the zero points of the transfer functions are suppressed 
for $\alpha \sim 5\times10^{14}$, even when there exist observational errors. 
As mentioned in a previous paper~\cite{KSY04}, 
this value arises from the difference between the relative magnitudes of 
the transfer functions of TT and EE due to the tight coupling. 

We have also investigated the possibility of 
constraining the cosmological parameters by using our inversion method, 
introducing the quantity $D$, 
which represents the difference between the reconstructed spectra 
for different values of $\alpha$. 
In the ideal case with no observational error, 
we have shown that $D$ takes its minimum value 
at the correct values of the cosmological parameters, 
independently of the shape of $P(k)$. 

Then, to determine how tightly we can constrain the cosmological parameters 
in a realistic situation, we have performed simulations 
by taking the PLANCK observational errors into account. 
We generated 1000 realizations 
and determined the values of the cosmological parameters 
which minimize the value of $D$ for each realization. 
By constructing histograms of these values, 
we estimated their probability distributions and $1\sigma$ errors. 
However, to save the computational time, 
we have performed only one- and two-dimensional analyses. 

In the one-dimensional analysis, 
where we vary only one of the cosmological parameters at a time, 
with the others fixed at the assumed values, 
we found that their probability distributions are nearly Gaussian, 
with the mean values close to the correct values. 
We also found that $\Omega_\Lambda$ is constrained 
most severely, with variances of a few percent relative to the mean values. 
This is because the variation of $\Omega_\Lambda$ is equivalent to 
the variation of $\Omega_K$ for fixed $\Omega_m$, 
which significantly affects the angular diameter distance to the LSS, 
and an incorrect choice of $\Omega_\Lambda$ displaces 
the locations of the singularities in such a way that the consistency of 
the shapes of the TT and EE spectra is lost. 
This leads to a large value of $D$. 

In the two-dimensional analysis, we have investigated two cases, 
one in which only $\Omega_b$ and $\Omega_m$ are varied, 
with $h$ and $\Omega_K$ fixed, 
and one in which only $h$ and $\Omega_\Lambda$ are varied, 
with $\Omega_bh^2$ and $\Omega_mh^2$ fixed. 
In the latter case, 
there is a degeneracy in the $(h\,,\,\Omega_\Lambda)$ plane, 
because the angular diameter distance to the LSS, 
which determines the scale of the spectrum, 
depends on both $h$ and $\Omega_\Lambda$~\cite{EB99}. 
In our analysis, 
the sparse mesh and narrow parameter range caused an artificial peak 
in the probability distribution. 
Thus, we need to improve our computational method 
by optimizing the numerical code. 

In any case, from the one-dimensional analysis, 
we may conclude that even if we allow an arbitrary functional form of $P(k)$, 
the cosmological parameters can be constrained by our inversion method, 
though this conclusion must be regarded as tentative, 
because we have explored only a limited range of the full, 
multi-dimensional parameter space in this paper. 

It should be mentioned that the constraints obtained with our method are, 
of course, weaker than 
those obtained with the conventional parameter-fitting method 
if the primordial spectrum $P(k)$ exhibits a simple power-law form, 
as assumed in the latter. 
As the observational accuracy increases, however, 
there is a good chance that we will observe nontrivial effects of 
fundamental quantum physics~\cite{MB01,BM01,MB03} 
and/or non-slow-roll evolution of 
the inflation-driving scalar field~\cite{AAS92,JY99,LL01,ACE01}. 
These effects may impart complicated features on the primordial power spectrum 
that are beyond a simple power law. 
If this is indeed the case, 
we would not be able to obtain accurate values of the cosmological parameters 
using the conventional method, 
which restricts the spectral shape from the beginning. 
Our method would serve as a powerful tool in such a situation 
and could be more appropriate for the next generation of 
high-precision experiments, such as PLANCK, which are intended to 
provide data from which we can derive 
not only accurate values of the cosmological parameters 
but also a more precise shape of the primordial spectrum. 

\renewcommand{\arraystretch}{1.5}

\begin{table}[t]
\begin{center}
\caption{Estimated values of the cosmological parameters 
obtained from the probability distributions shown in Fig.~\ref{HPARA1}. 
\label{VPARA1}}
\vspace{1cm}
\begin{tabular}{c|c|c|c}
\hline\hline
& Assumed value & Estimated value & Fixed parameters \\
\hline
$h$  & $0.70$  & $0.703^{+0.027}_{-0.025}$
   & $\Omega_bh^2$, $\Omega_mh^2$, $\Omega_K$, $\tau$ \\
\hline
$\Omega_b$ & $0.050$ & $0.0491^{+0.0031}_{-0.0027}$ 
   & $h$, $\Omega_mh^2$, $\Omega_K$, $\tau$ \\
\hline
$\Omega_m$ & $0.30$  & $0.293^{+0.020}_{-0.017}$
   & $h$, $\Omega_bh^2$, $\Omega_K$, $\tau$ \\
\hline
$\Omega_\Lambda$ & $0.70$  & $0.701^{+0.011}_{-0.011}$
   & $h$, $\Omega_bh^2$, $\Omega_mh^2$, $\tau$ \\
\hline
\end{tabular}
\end{center}
\end{table}

\begin{table}[t]
\begin{center}
\caption{Estimated parameters 
obtained from the probability distributions shown in Fig.~\ref{HBM2}. 
Note that the range of $\Omega_b$ we investigated may not be large enough, 
as seen in Fig.~\ref{HBM2}. 
\label{VBM2}}
\vspace{1cm}
\begin{tabular}{c|c|c|c}
\hline\hline
& Assumed value & Estimated value & Fixed parameters \\
\hline
$\Omega_b$ & $0.050$ & $0.0465^{+0.0037}_{-0.0036}$ 
   & $h$, $\Omega_K$, $\tau$ \\
\hline
$\Omega_m$ & $0.30$ & $0.282^{+0.030}_{-0.028}$ 
   & $h$, $\Omega_K$, $\tau$ \\
\hline
\end{tabular}
\end{center}
\end{table}

\section*{Acknowledgements}
We would like to thank Makoto Matsumiya for discussions. 
This work was supported in part by JSPS Grants-in-Aid for 
Scientific Research (12640269 for M.~S. and 16340076 for J.~Y.), 
by a Monbu-Kagakusho Grant-in-Aid for Scientific Research (S) 
(14102004 for M.~S.), 
and by a Grant-in-Aid for the 21st Century COE 
``Center for Diversity and Universality in Physics" from 
the Ministry of Education, Culture, Sports, Science and Technology (MEXT) 
of Japan. 
N.~K. is supported by Research Fellowships of JSPS 
for Young Scientists (04249). 

\appendix

\section{Derivation of Inversion Formula} \label{DERIVATION}

Here we review our method of the reconstruction of $P(k)$ 
from both the CMB temperature and polarization spectra 
for scalar-type perturbations~\cite{MSY02,MSY03,KSY04}. 
We assume Gaussian and adiabatic primordial fluctuations. 
Although we restrict our discussions to a flat universe, 
it is easy to extend our formalism to a nonflat universe 
as mentioned in Sec.~\ref{FORMULA}. 

The angular power spectrum of the CMB anisotropy is expressed as 
\begin{eqnarray}
\frac{2\ell+1}{4\pi} C^{X\bar{X}}_\ell
=\frac{1}{2\pi^2} \int_0^\infty \! \frac{dk}{k} \,
 \frac{k^3 \bigl\langle X_\ell^*(\eta_0,k)
 \bar{X}_\ell(\eta_0,k) \bigl\rangle}{2\ell+1},
\label{CL}
\end{eqnarray}
where $X_\ell(\eta,k)$ and $\bar{X}_\ell(\eta,k)$ 
are either $\Theta_\ell(\eta,k)$ and $E_\ell(\eta,k)$ 
which represent multipole moments of temperature fluctuations 
and E-mode polarization in Fourier space, respectively.  
$k$ is the comoving wavenumber and $\eta$ is the conformal time 
with $\eta_0$ being the present value. 
These are expressed in the integral form of the Boltzmann equations 
($\ell \ge 2$)~\cite{SZ96}: 
\begin{eqnarray}
\frac{\Theta_\ell(\eta_0,k)}{2\ell+1}
&=& \int_0^{\eta_0} \! d\eta \, \bigg\{
    \left[ (\Theta_0+\Psi) {\cal V}(\eta)
    +(\dot{\Psi}-\dot{\Phi}) e^{-\tau(\eta)}
    \right] j_\ell(k\Delta\eta) \nonumber \\
& & {}+V_b {\cal V}(\eta) j'_\ell(k\Delta\eta)
    +\frac{1}{2} \Pi_2 {\cal V}(\eta)
    \left[ 3j''_\ell(k\Delta\eta)+j_\ell(k\Delta\eta) \right]
    \bigg\},
\label{TL} \\
\frac{E_\ell(\eta_0,k)}{2\ell+1}
&=& -\frac{3}{2} \sqrt{\frac{(\ell+2)!}{(\ell-2)!}}
    \int_0^{\eta_0} \! d\eta \,
    \Pi_2 {\cal V}(\eta)\,
    \frac{j_\ell(k\Delta\eta)}{(k\Delta\eta)^2},
\label{EL}
\end{eqnarray}
where $\Pi_2 \equiv (\Theta_2-\sqrt{6} E_2)/10$, 
$\Delta\eta \equiv \eta_0-\eta$, 
and the overdot denotes a derivative with respect to the conformal time. 
Here $V_b$ is the baryon fluid velocity, 
$\Psi$ and $\Phi$ are the Newton potential 
and the spatial curvature perturbation in the Newton gauge, 
respectively~\cite{KS84}, and 
\begin{eqnarray}
{\cal V}(\eta) \equiv \dot{\tau} e^{-\tau(\eta)}, \quad
    \tau(\eta) \equiv \int_{\eta}^{\eta_0} \! \dot{\tau} d\eta,
\label{VIS&TAU}
\end{eqnarray}
are the visibility function and the optical depth for Thomson scattering, 
respectively. 
In the limit that 
the thickness of the last scattering surface (LSS) is negligible, we have 
${\cal V}(\eta) \approx \delta(\eta-\eta_*)$ and
$e^{-\tau(\eta)} \approx \theta(\eta-\eta_*)$, 
where $\eta_*$ is the recombination time 
when the visibility function is maximum~\cite{HS95}. 
To obtain a better approximation, 
we take into account the thickness of the LSS. 
This is required especially for the polarization, 
since the CMB polarization is mainly generated 
within the thickness of the LSS. 
The approximation is to neglect the oscillations of 
the spherical Bessel functions in the integrals. 
Applying this approximation to Eqs.~(\ref{TL}) and (\ref{EL}), we have 
\begin{eqnarray}
\frac{\Theta_\ell(\eta_0,k)}{2\ell+1} &\approx&
 \left\{ \int_{\eta_{*{\rm start}}}^{\eta_{*{\rm end}}} \! d\eta \,
         \left[ (\Theta_0+\Psi) {\cal V}(\eta)
         +(\dot{\Psi}-\dot{\Phi}) e^{-\tau(\eta)} \right] \right\} j_\ell(kd)
         \nonumber \\ & &{}
+\left\{ \int_{\eta_{*{\rm start}}}^{\eta_{*{\rm end}}} \! d\eta \,
         \Theta_1(\eta,k) {\cal V}(\eta) \right\} j'_\ell(kd)
\equiv\frac{\Theta^{\rm app}_\ell(\eta_0,k)}{2\ell+1},
\label{TLAPP} \\
\frac{E_\ell(\eta_0,k)}{2\ell+1} &\approx&
 \sqrt{\frac{(\ell+2)!}{(\ell-2)!}}
 \left\{ -\frac{3}{2} 
 \int_{\eta_{*{\rm start}}}^{\eta_{*{\rm end}}} \! d\eta \,
 \frac{\Pi_2}{(k\Delta\eta)^2} {\cal V}(\eta) \right\} j_\ell(kd)
\equiv\frac{E^{\rm app}_\ell(\eta_0,k)}{2\ell+1},
\label{ELAPP}
\end{eqnarray}
where $d \equiv \eta_0-\eta_*$ is the conformal distance 
from the present to the LSS 
and $\eta_{*{\rm start}}$ and $\eta_{*{\rm end}}$ are the times 
when the recombination starts and ends, respectively. 
We have also replaced $V_b$ by $\Theta_1$ 
and neglected the quadrupole term in Eq.~(\ref{TL}), 
by adopting the tight coupling approximation~\cite{HS95}. 
We define time-integrated transfer functions 
within the thickness of the LSS, $f(k)$, $g(k)$, and $h(k)$ as 
\begin{eqnarray}
\int_{\eta_{*{\rm start}}}^{\eta_{*{\rm end}}} \! d\eta \,
\left[ (\Theta_0+\Psi)(\eta,k){\cal V}(\eta)
      +(\dot{\Psi}-\dot{\Phi})(\eta,k) e^{-\tau(\eta)} \right]
&\equiv& f(k) \Phi(0,\bm{k}),
\label{TRANSF} \\
\int_{\eta_{*{\rm start}}}^{\eta_{*{\rm end}}} \! d\eta \,
\Theta_1(\eta,k){\cal V}(\eta)
&\equiv& g(k) \Phi(0,\bm{k}),
\label{TRANSG} \\
-\frac{3}{2} \int_{\eta_{*{\rm start}}}^{\eta_{*{\rm end}}} \! d\eta \,
 \frac{\Pi_2(\eta,k)}{(k(\eta_0-\eta))^2} {\cal V}(\eta)
&\equiv& h(k) \Phi(0,\bm{k}).
\label{TRANSH}
\end{eqnarray}
Here we have separated terms of the transfer functions 
which are dependent only on the cosmological parameters, 
and the primordial curvature perturbation which leads 
the primordial power spectrum is defined as 
\begin{eqnarray}
P(k) \equiv \langle |\Phi(0,\bm{k})|^2 \rangle.
\label{PK}
\end{eqnarray}
Let us compare the magnitudes of the transfer functions. 
At the recombination the quadrupole, dipole, and monopole are related as 
\begin{eqnarray}
\Theta_1
&\sim& \frac{k}{a_* H_*}(\Theta_0+\Psi)
 \sim  k\eta_*(\Theta_0+\Psi)
 \sim  kd \left(\frac{\eta_*}{\eta_0}\right) (\Theta_0+\Psi),
\label{DIPOLE} \\
\frac{\Pi_2}{(kd)^2}
&\sim& \frac{1}{(kd)^2} \frac{k}{a_* n_e \sigma_T} \Theta_1
 \sim  \frac{k\eta_*}{(kd)^2} \frac{H_*}{n_e \sigma_T} \Theta_1
 \sim  \frac{1}{kd} \left(\frac{\eta_*}{\eta_0}\right)
       \left(\frac{H_*}{n_e \sigma_T}\right) \Theta_1,
\label{QUADRUPOLE}
\end{eqnarray}
where $n_e$ is the number density of free electrons, 
$\sigma_T$ is the cross section of the Thomson scattering, 
and the subscript $*$ denotes the value at the recombination. 
Since $\eta_*/\eta_0 \sim 0.02$ 
and the mean free time of the Thomson scattering is much shorter than 
the cosmic expansion time, $H_*/(n_e \sigma_T) \sim 10^{-3}$~\cite{KS84}, 
$\Theta_1 \sim 10^{-2} (kd)(\Theta_0+\Psi)$ 
and $\Pi_2/(kd)^2 \sim 10^{-5} (kd)^{-1}\Theta_1$. 
Thus, at $kd \sim \ell \sim O(10^2)$, $f(k) \sim g(k) \sim 10^7 h(k)$. 

Substituting Eqs.~(\ref{TLAPP}) and (\ref{ELAPP}) into Eq.~(\ref{CL}), 
we obtain the approximated TT, EE, and TE angular power spectra, 
\begin{eqnarray}
\frac{2\ell+1}{4\pi} C^{TT,\,{\rm app}}_\ell
&=& \frac{2\ell+1}{2\pi^2} \int_0^\infty \! \frac{dk}{k} \, k^3 P(k)
     \left[ f(k) j_\ell(kd)+g(k) j'_\ell(kd) \right]^2,
\label{CLTTAPP} \\
\frac{2\ell+1}{4\pi} C^{EE,\,{\rm app}}_\ell
&=& \frac{2\ell+1}{2\pi^2} \,\frac{(\ell+2)!}{(\ell-2)!} \,
 \int_0^\infty \! \frac{dk}{k} \, k^3 P(k) 
    \left[ h(k) j_\ell(kd) \right]^2,
\label{CLEEAPP} \\
\frac{2\ell+1}{4\pi} C^{TE,\,{\rm app}}_\ell
&=& \frac{2\ell+1}{2\pi^2}\sqrt{\frac{(\ell+2)!}{(\ell-2)!}}
\nonumber \\
& & \!\!\times \int_0^\infty \! \frac{dk}{k} \, k^3 P(k) 
\left[ f(k) j_\ell(kd)+g(k) j'_\ell(kd) \right] h(k) j_\ell(kd).
\label{CLTEAPP}
\end{eqnarray}
The angular correlation function of the CMB temperature fluctuations 
is defined as 
\begin{eqnarray}
C^{TT}(\theta)
\equiv \left\langle \Theta(\hat{\bm{n}}_1)
       \Theta(\hat{\bm{n}}_2) \right\rangle
     = \sum_{\ell=0}^{\infty} \frac{2\ell+1}{4\pi}
       C^{TT}_\ell P_\ell(\cos\theta),
\quad \cos\theta=\hat{\bm{n}}_1 \cdot \hat{\bm{n}}_2 \, .
\label{CRTT}
\end{eqnarray}
Similarly, we introduce the following quantities for the polarization: 
\begin{eqnarray}
\tilde{C}^{EE}(\theta)
&\equiv& \sum_{\ell=0}^{\infty} \frac{2\ell+1}{4\pi}
         \frac{(\ell-2)!}{(\ell+2)!} \, C^{EE}_\ell P_\ell(\cos\theta),
\label{CREE} \\
\tilde{C}^{TE}(\theta)
&\equiv& \sum_{\ell=0}^{\infty} \frac{2\ell+1}{4\pi}
         \sqrt{\frac{(\ell-2)!}{(\ell+2)!}} \, C^{TE}_\ell P_\ell(\cos\theta),
\label{CRTE}
\end{eqnarray}
which are not the conventional angular correlation functions 
but they turn out to be convenient for inversion. 
Here we introduce a new variable $r$ instead of $\theta$ defined as 
\begin{eqnarray}
r=2d \sin \frac{\theta}{2},
\label{CRAPP}
\end{eqnarray}
which is the conformal distance between two points on the LSS. 
From now on, we focus only on small angular scales, 
corresponding to $r \ll d$, which is valid where $\ell \gtrsim O(10)$. 
For the TT spectrum, substituting Eq.~(\ref{CLTTAPP}) into Eq.~(\ref{CRTT}), 
we can derive formulas for the approximated angular correlation functions 
in terms of $P(k)$. 
With the help of the Fourier sine formula, 
we obtain a first-order differential equation for $P(k)$, 
\begin{eqnarray}
&&-k^2f^2(k)P'(k)+\left[ -2k^2f(k)f'(k)+kg^2(k) \right] P(k)
\nonumber \\
&&=4\pi \int_0^\infty \! dr \,
   \frac{1}{r} \frac{\partial}{\partial r} 
   \{ r^3 C^{TT,\,{\rm app}}(r) \} \sin kr
   \equiv S^{TT}(k).
\label{FORMULATT}
\end{eqnarray}
This is the basic equation for the inversion 
of the TT angular power spectrum 
to the primordial curvature perturbation spectrum. 
Although the above differential equation is singular at $f(k)=0$ 
since the transfer functions are oscillatory, 
we can find the values of $P(k)$ at such points, say $k=k_s$, as 
\begin{eqnarray}
P(k_s)=\frac{S^{TT}(k_s)}{k_s \, g^2(k_s)} \quad {\rm for} \quad f(k_s)=0,
\label{BCTT}
\end{eqnarray}
assuming that $P'(k)$ is finite at $k=k_s$. 
We can then solve Eq.~(\ref{FORMULATT}) 
as a boundary value problem between the singularities. 
Similarly, for the EE and TE spectrum, 
substituting Eqs.~(\ref{CLEEAPP}) and (\ref{CLTEAPP}) 
into Eqs.~(\ref{CREE}) and (\ref{CRTE}), respectively, and so on, 
we obtain the algebraic equations for $P(k)$, 
\begin{eqnarray}
kh^2(k)P(k)
&=& 4\pi \int_0^\infty \! dr \, r\tilde{C}^{EE,\,{\rm app}}(r) \sin kr
    \equiv S^{EE}(k),
\label{FORMULAEE} \\
kf(k)h(k)P(k)
&=& 4\pi \int_0^\infty \! dr \, r\tilde{C}^{TE,\,{\rm app}}(r) \sin kr
    \equiv S^{TE}(k).
\label{FORMULATE}
\end{eqnarray}
In this case, we can find $P(k)$ 
except for the singularities of $h(k)=0$ for EE, 
and those of $f(k)=0$ and $h(k)=0$ for TE, respectively. 
These are the basic inversion formulas for the EE and TE angular power spectra.
If we use inversion formulas (\ref{FORMULATT}), (\ref{FORMULAEE}) 
and (\ref{FORMULATE}) separately, 
the reconstructed $P(k)$ suffers from large numerical errors 
around the singularities of the respective equations. 
However, the fact that the zero points of $f(k)$ and $h(k)$ are different 
from each other can be used to resolve this numerical problem as follows. 
We construct a combined inversion formula 
by multiplying Eq.~(\ref{FORMULAEE}) by some factor $\alpha$ 
and adding it to Eq.~(\ref{FORMULATT}),
\begin{eqnarray}
&&-k^2f^2(k)P'(k)+\left[ -2k^2f(k)f'(k)+kg^2(k)+\alpha kh^2(k) \right] P(k)
\nonumber \\
&&=S^{TT}(k)+\alpha S^{EE}(k).
\label{FORMULACOM}
\end{eqnarray}
The boundary conditions are similarly given 
by the values of $P(k)$ at zero points of $f(k)$ as 
\begin{eqnarray}
P(k_s)=\frac{S^{TT}(k_s)+\alpha S^{EE}(k_s)}
            {k_s \left[ g^2(k_s)+\alpha h^2(k_s) \right]}
\quad {\rm for} \quad f(k_s)=0,
\label{BCCOM}
\end{eqnarray}
assuming that $P'(k)$ is finite at $k=k_s$. 
Here we have taken $\alpha$ to be independent of $k$ for simplicity 
and this free parameter $\alpha$ controls 
the contribution of EE relative to TT. 
If we take an appropriate value of $\alpha$ 
so that the contribution of EE is comparable to that of TT, 
the solution of Eq.~(\ref{FORMULACOM}) becomes numerically stable 
even around the singularities. 
This is because the contribution of EE dominates 
near the singularities of TT given by $f(k)=0$, and vice versa. 
We have found such an appropriate value of $\alpha$ 
is $\sim 10^{13}-10^{15}$ if we assume no observational error~\cite{KSY04}. 
The origin of this number 
comes from the fact that the transfer function $h(k)$ for EE 
is intrinsically smaller than the transfer functions $f(k)$ and $g(k)$ 
for TT by a factor $\sim 10^{-7}$ as explained above 
and their squares are contained 
in the left-hand side of Eq.~(\ref{FORMULACOM}). 
Since the TE formula, Eq.~(\ref{FORMULATE}), 
which is singular not only at $h(k)=0$ but also at $f(k)=0$, 
is difficult to handle, we do not use it so far. 

For either TT or EE, 
the approximated spectrum $C^{XX,\,{\rm app}}_\ell$ 
($X=T$ or $E$) expressed as Eq.~(\ref{CLTTAPP}) or (\ref{CLEEAPP}) 
has relative errors as large as about $20-30\%$ 
to the exact one $C^{XX,\,{\rm ex}}_\ell$ expressed as Eq.~(\ref{CL}). 
Unless we correct such errors due to the approximation, 
it leads wrong $P(k)$ when solving Eq.~(\ref{FORMULACOM}) 
which uses the observed spectrum $C^{XX,\,{\rm obs}}_\ell$. 
Therefore, we introduce the ratio, 
\begin{eqnarray}
b^{XX}_\ell \equiv
\frac{C^{XX,\,{\rm ex}}_\ell}{C^{XX,\,{\rm app}}_\ell}\,.
\label{RATIO}
\end{eqnarray}
This ratio is found to be almost independent of $P(k)$. 
We explain the reason as follows. 
In both $C^{XX,\,{\rm ex}}_\ell$ and $C^{XX,\,{\rm app}}_\ell$, 
the transfer functions including the spherical Bessel functions 
are rapidly oscillating functions compared to a reasonable $P(k)$ 
for sufficiently large $\ell$. 
Thus, when we take the ratio, the dependence on $P(k)$ 
effectively cancels out. 
We use this property of $b^{XX}_\ell$ 
with some fiducial primordial spectrum $P^{(0)}(k)$ 
such as a scale-invariant one. 
That is, we first calculate 
$b^{XX,\,(0)}_\ell=C^{XX,\,{\rm ex}(0)}_\ell/C^{XX,\,{\rm app}(0)}_\ell$, 
from $P^{(0)}(k)$, 
and then we use $C^{XX,\,{\rm obs}}_\ell/b^{XX,\,(0)}_\ell$ 
which gives a better guess to $C^{XX,\,{\rm app}}_\ell$, 
instead of $C^{XX,\,{\rm obs}}_\ell$, 
in the right-hand side of Eq.~(\ref{FORMULACOM}). 
As a result, we obtain $P(k)$ with much better accuracy. 

Finally, substituting Eqs.~(\ref{CRTT}) and (\ref{CREE}) 
into Eq.~(\ref{FORMULACOM}) and replacing $C^{XX,\,{\rm app}}_\ell$ 
by $C^{XX,\,{\rm obs}}_\ell/b^{XX,\,(0)}_\ell$, 
we obtain the inversion formula (\ref{INVERSIONCOM}) in Sec.~\ref{FORMULA}.

\end{document}